\documentclass[a4paper,10pt]{article}
\pdfoutput=1
\usepackage{graphicx,rotating,hyperref,slashed,amsmath,xcolor,amssymb,amsfonts,colortbl,cite} 
\makeatletter
\hypersetup{colorlinks,bookmarksopen,bookmarksnumbered,
linkcolor=blus,pdfstartview=FitH,urlcolor=rossos,citecolor=verde}
\allowdisplaybreaks	 
\numberwithin{equation}{section} 

\hypersetup{%
    ,urlcolor=black 
    ,citecolor=black
    ,linkcolor=black
    }
\usepackage{mciteplus}

\AtBeginDocument{
  \hypersetup{
    urlcolor=blue,
    citecolor=blue,
    linkcolor=black,
  }%
}

\def\bea{\begin{eqnarray}}
\def\eea{\end{eqnarray}}

\def\lsim{\mathrel{\rlap{\lower3pt\hbox{\hskip0pt$\sim$}}
   \raise1pt\hbox{$<$}}}         
\def\gsim{\mathrel{\rlap{\lower4pt\hbox{\hskip1pt$\sim$}}
   \raise1pt\hbox{$>$}}}         

 \newcommand{\sfootnote}[1]{}
\definecolor{bluc}{cmyk}{1,1,0,0.1}
\definecolor{rossoCP3}{cmyk}{0,.88,.77,.40}
\definecolor{rosso}{cmyk}{0,1,1,0.4}
\definecolor{rossos}{cmyk}{0,1,1,0.55}
\definecolor{rossoc}{cmyk}{0,1,1,0.2}
\definecolor{verdes}{cmyk}{0.92,0,0.59,0.4}

\hypersetup{colorlinks, bookmarksopen, bookmarksnumbered,
citecolor=verdes, linkcolor=bluc, pdfstartview=FitH, urlcolor=rossos}

\usepackage{multicol}
\usepackage{color}
\definecolor{rosso}{cmyk}{0,1,1,0.4}
\definecolor{rossos}{cmyk}{0,1,1,0.55}
\definecolor{rossoc}{cmyk}{0,1,1,0.2}
\definecolor{blu}{cmyk}{1,1,0,0.3}
\definecolor{blus}{cmyk}{1,1,0,0.6}
\definecolor{bluc}{cmyk}{1,1,0,0.1}
\definecolor{verde}{cmyk}{0.92,0,0.59,0.25}
\definecolor{verdec}{cmyk}{0.92,0,0.59,0.15}
\definecolor{verdes}{cmyk}{0.92,0,0.59,0.4}

\skip\@mpfootins = \skip\footins
\renewcommand\&{&}

\def\circa#1{\,\raise.3ex\hbox{$#1$\kern-.75em\lower1ex\hbox{$\sim$}}\,}

\newcommand{\be}{\begin{equation}}
\newcommand{\ee}{\end{equation}}
\newfam\rsfsfam
\def\mathscr#1{{\fam\rsfsfam\relax#1}}

\def\circa#1{\,\raise.3ex\hbox{$#1$\kern-.75em\lower1ex\hbox{$\sim$}}\,}
\makeatletter

\def\hhref#1{\href{http://arxiv.org/abs/#1}{arXiv:#1}} 

\newcommand{\doi}[1]{\href{http://dx.doi.org/#1}{[doi]}}

\def\hhref#1{\href{http://arxiv.org/abs/#1}{arXiv:#1}} 
 
\def\art{\@ifnextchar[{\eart}{\oart}}
\def\eart[#1]#2#3#4#5#6{{\rm #2}, {\em #3 \bf #4} {\rm (#6) #5} ({\em #1})}

\def\article{\@ifnextchar[{\earticle}{\oarticle}}
\def\oarticle#1#2#3#4#5#6{{\rm #1}, {\em ``#6''}, {\rm #2 #3 (#5) #4}}
\def\earticle[#1]#2#3#4#5#6#7{{\rm #2}, {\em ``#7''}, {\rm #3 #4 (#6) #5}  [\hhref{#1}]}
\def\hepart[#1]#2{{\rm #2, \em#1}}
\def\heparticle[#1]#2#3{#2, {\em ``#3''} [\hhref{#1}]}

\newcounter{alphaequation}[equation]
\def\thealphaequation{\theequation\hbox to
0.6em{\hfil\alph{alphaequation}\hfil}}
\def\eqnsystem#1{
\def\@eqnnum{{\rm (\thealphaequation)}}
\def\@@eqncr{\let\@tempa\relax \ifcase\@eqcnt \def\@tempa{& & &} \or
  \def\@tempa{& &}\or \def\@tempa{&}\fi\@tempa
  \if@eqnsw\@eqnnum\refstepcounter{alphaequation}\fi
\global\@eqnswtrue\global\@eqcnt=0\cr}
\refstepcounter{equation} \let\@currentlabel\theequation \def\@tempb{#1}
\ifx\@tempb\empty\else\label{#1}\fi
\refstepcounter{alphaequation}
\let\@currentlabel\thealphaequation
\global\@eqnswtrue\global\@eqcnt=0 \tabskip\@centering\let\\=\@eqncr
$$\halign to \displaywidth\bgroup \@eqnsel\hskip\@centering
$\displaystyle\tabskip\z@{##}$&\global\@eqcnt\@ne
\hskip2\arraycolsep\hfil${##}$\hfil& \global\@eqcnt\tw@\hskip2\arraycolsep
$\displaystyle\tabskip\z@{##}$\hfil
\tabskip\@centering&\llap{##}\tabskip\z@\cr}
\def\endeqnsystem{\@@eqncr\egroup$$\global\@ignoretrue} \makeatother

\definecolor{fiorentina}{rgb}{.5,0,.5}

\begin{document}



\setcounter{page}{1} \baselineskip=15.5pt \thispagestyle{empty}

 
\vspace{0.8cm}
\begin{center}

{\fontsize{19}{28}\selectfont  \sffamily \bfseries {Kinematic anisotropies and pulsar timing arrays}}

\end{center}

\vspace{0.2cm}

\begin{center}
{\fontsize{13}{30}\selectfont  Gianmassimo Tasinato$^{1,2}$ } 
\end{center}

\begin{center}

\vskip 8pt
\textsl{$^{1}$ Dipartimento di Fisica e Astronomia, Universit\`a di Bologna,  Italia}
\\
\textsl{$^{2}$ Physics Department, Swansea University, SA28PP, United Kingdom}\\
\textsl{\texttt{email}: g.tasinato2208 at gmail.com }\\
\vskip 7pt

\end{center}

\smallskip
\begin{abstract}
\noindent 
Doppler anisotropies, induced by our relative motion with respect to the source rest frame, are a guaranteed property of stochastic gravitational wave backgrounds of cosmological origin. If detected by future pulsar timing array  measurements,   they will provide interesting information on the physics sourcing gravitational waves,  which is hard or even impossible to extract  from measurements of  the isotropic part of the background only. 
 We analytically determine the pulsar response function to  kinematic anisotropies, including possible effects due to parity violation, to features in the frequency dependence of the isotropic part of the spectrum, as well as to the  presence of  extra scalar and vector polarizations. For the first time, we show  how the  sensitivity to  different  effects   crucially depends on the pulsar configuration   with respect to  the relative  motion among frames.   Correspondingly,  we propose examples of strategies of detection,  each aimed at   exploiting future measurements of kinematic anisotropies for characterizing  distinct features of the cosmological gravitational wave background.
\end{abstract}


\section{Introduction}
Four 
 pulsar timing array (PTA) collaborations \cite{NANOGrav:2023gor,Reardon:2023gzh,EPTA:2023fyk,Xu:2023wog} 
recently detected
a signal compatible with a stochastic gravitational
wave background (SGWB). At the moment,
 various open questions remain on the  origin of the gravitational wave (GW)
signal, since diverse sources, from astrophysical to cosmological,
are compatible with current observations (see e.g. \cite{Ellis:2023oxs} for  a recent multi-model
assessment). Properties of the SGWB can in principle allow us to distinguish
astrophysical from cosmological sources of GW: parity violation \cite{Jackiw:2003pm,Alexander:2004us,Lue:1998mq,Satoh:2007gn,Contaldi:2008yz,Anber:2009ua,Alexander:2009tp,Takahashi:2009wc,Anber:2012du,Bartolo:2016ami,Mylova:2019jrj,Ozsoy:2021onx,Fu:2023aab}, non-Gaussianities (see e.g. \cite{Bartolo:2018qqn,Bartolo:2018evs,Powell:2019kid,Tasinato:2022xyq}),
or anisotropies (see \cite{LISACosmologyWorkingGroup:2022kbp} for a recent survey). If the SGWB has cosmological
origin, besides its intrinsic anisotropies, the signal is expected
to be characterised by a Doppler anisotropy due to our relative motion with respect
to the source rest frame. As found by cosmic microwave background (CMB) experiments \cite{Smoot:1977bs,Kogut:1993ag,WMAP:2003ivt,Planck:2013kqc}, our
velocity with respect to the cosmic rest frame has an amplitude of size $\beta \,=\, v/c \,=\,1.23 \times 10^{-3}$ with respect to
 the velocity of light, and points in the direction $(l, b)\,=\,(264^{\rm o}, 48^{\rm o})$. Correspondigly, the size of the  dipolar kinematic anisotropy of a cosmological SGWB should be one-thousandth smaller than
 the amplitude of the isotropic part of the GW spectrum. This is smaller in size than the expected anisotropies
 of the astrophysical SGWB in the PTA band (see e.g. \cite{Cornish:2013aba,Cornish:2015ikx,Taylor:2020zpk,Becsy:2022pnr,Allen:2022dzg} and in particular \cite{Sato-Polito:2021efu} for a recent study). But, as for the CMB, it is potentially well larger than intrinsic cosmological anisotropies (see, e.g., the studies  \cite{Alba:2015cms,Contaldi:2016koz,Geller:2018mwu,Bartolo:2019oiq,Bartolo:2019zvb,Bartolo:2019yeu,Dimastrogiovanni:2021mfs}). Kinematic anisotropies is a topic of active research in the context of GW interferometers \cite{LISACosmologyWorkingGroup:2022kbp,Cusin:2022cbb,Bertacca:2019fnt,ValbusaDallArmi:2022htu,Chung:2022xhv,Chowdhury:2022pnv}. It
 is then worth 
 asking what   information we can obtain from a  possible future detection of kinematic anisotropies with PTA experiments.
 
 \smallskip
 
 This is the scope of this work. The development and refinement of detection techniques of anisotropies
 of SGWB from PTA observations have a long history -- see e.g. \cite{Anholm:2008wy,Mingarelli:2013dsa,Taylor:2013esa,Gair:2014rwa,Cornish:2014rva,Mingarelli:2017fbe,Ali-Haimoud:2020ozu,Ali-Haimoud:2020iyz}  -- often
 borrowing and elaborating techniques developed in the context
  of ground-based \cite{Allen:1996gp,Ballmer:2005uw,Thrane:2009fp,Renzini:2018vkx,Payne:2020pmc} 
and space-based \cite{Cornish:2001hg,Baker:2019ync,Banagiri:2021ovv,Contaldi:2020rht,LISACosmologyWorkingGroup:2022jok}  interferometers. Current PTA measurements 
do not find yet indications of SGWB anisotropies \cite{Taylor:2015udp,NANOGrav:2023tcn}, but as data are becoming more and more accurate, a detection might be forthcoming (see
also \cite{Chu:2021krj,Bernardo:2022rif,Bernardo:2023mxc} for recent studies of related aspects of GW observations with PTA).  Since  in this work we specialize
 to kinematic anisotropies, we  express  our findings
 in the most 
  convenient way to extract information from a detection of Doppler effects
  in the SGWB. We stress the geometrical aspects of the problem, and we  
   make manifest how the  sensitivity to  different GW effects depends on the position of pulsars with respect to the 
  relative velocity vector among frames. These findings -- of which many are new while others are known but we set them in a new perspective -- 
   can be useful to plan future PTA observations. In fact,  as we can plan the location
   of ground based interferometers in the Earth in order to extract most physics
   from observations (for example, see the recent study \cite{Branchesi:2023mws} in the context of the Einstein Telescope), we can also plan which pulsars are worth monitoring to learn new physics
   from GW measurements. This is an invaluable opportunity for 
   forthcoming  observations with SKA facilities, which will monitor and detect signals from a large number of pulsars \cite{Keane:2014vja} with important opportunities for GW physics: see e.g. \cite{Janssen:2014dka}. 
    The physical effects we
  explore using SGWB kinematic
  anisotropies in PTA band are:

 \begin{itemize} 
 \item[$\triangleright$] {\bf Parity violation:} Well motivated
early-universe scenarios predict the existence of parity-violating effects in
gravitational and gauge-field
interactions, with interesting cosmological and GW consequences, see e.g. \cite{Jackiw:2003pm,Alexander:2004us,Lue:1998mq,Satoh:2007gn,Contaldi:2008yz,Anber:2009ua,Alexander:2009tp,Takahashi:2009wc,Anber:2012du,Bartolo:2016ami,Mylova:2019jrj,Ozsoy:2021onx,Fu:2023aab}. These effects manifest as circular polarization
of the SGWB. This quantity is hard or even impossible to detect with single interferometers, when
focussing on the isotropic part of the background only: see
e.g. the clear analysis of  \cite{Smith:2016jqs}. Opportunities of detection arise when cross
correlating different experiments, see e.g. \cite{Crowder:2012ik,Seto:2020mfd,Orlando:2020oko},
or by probing anisotropies of the SGWB: \cite{Seto:2008sr,Seto:2006hf,Seto:2006dz,Domcke:2019zls}. 
 For the case of PTA measurements,  \cite{Kato:2015bye}
 shown for the first time that one needs to measure SGWB anisotropies
 for being sensitive to circular polarization (see also \cite{Belgacem:2020nda}).
 In our work, we show that kinematic anisotropies are in fact potentially able to probe
 circular polarization with PTA detections, and the PTA sensitivity depends on  the pulsar
 location with respect to the relative motion along frames. Hence, a careful plan of the pulsars
 to monitor is needed for increasing opportunities to detection (see sections \ref{secsetup}, \ref{secptar}, and \ref{sec_strate}).

\item[$\triangleright$]
{\bf  Frequency features in GW spectrum:}  Kinematic
anisotropies are obtained from Doppler boosting the isotropic part of the SGWB: they depend
on derivatives along frequency of this quantity \cite{Cusin:2022cbb}. Hence, Doppler
effects can be a complementary probe of the frequency dependence of the SGWB, alternative
to more direct methods (see e.g. \cite{Caprini:2019pxz}). Kinematic
anisotropies are more pronounced if the SGWB has features in frequency: suitable
combinations of PTA anisotropy measurements allow us to specifically detect the slope
of the background (see sections  \ref{secptar}, and \ref{sec_strate}).

\item[$\triangleright$]
{\bf  Presence of extra scalar and vector polarizations:}  Alternative
theories of gravity predict the existence of additional scalar and vector
polarizations with respect to the spin-2 ones of General Relativity (see e.g. \cite{Nishizawa:2009bf}
for a general discussion, and \cite{Romano:2016dpx} for a comprehensive review). Previous 
works investigated how PTA measurements can probe the GW polarization content, see e.g. \cite{Lee_2008,Gair:2012nm,Chamberlin:2011ev,Shao:2014wja,Gair:2015hra,Hotinli:2019tpc,Liu:2022skj}. Here we point out
that kinematic anisotropies, which depend on the slope of the spectrum, can be sensitive
to extra polarizations also when the latter give a small contribution to the isotropic
part of the spectrum (see sections  \ref{sec_MG}, and \ref{sec_strate}).

\end{itemize}
We discuss the  motivations and theoretical formulation of these topics  in sections \ref{secsetup},  \ref{sec_MG} and \ref{secptar}, while in section \ref{sec_strate} we   elaborate strategies of detection
of the  effects listed above. Two  Appendixes develop technical tools needed in the main text.

\section{Set-up}
\label{secsetup}

In this section we investigate the response of a PTA experiment
to an anisotropic SGWB characterized  by the GW intensity and GW circular polarization.
 We will   start appreciating how the PTA response  to  GW depend on the pulsar  configuration.
 In the next sections, 
the resulting 
formulas will then be applied  to the specific case of kinematic anisotropies.

\smallskip

The GW is expressed in terms of   fluctuations of the Minkowski metric
\be
d s^2\,=\,-d t^2+\left( \delta_{ij}+h_{ij}(t, \vec x)\right)\,d x^i d x^j
\,.
\ee
We decompose the  GW 
 in Fourier modes as
\be
\label{fouhij}
h_{ij}(t,\vec x)\,=\,\sum_{\lambda}\,\int_{-\infty}^{+\infty} d f\,\int d^2 \hat n\,{ e}^{-2 \pi i f \,\hat n \vec x}\,e^{2 \pi i f t}\,
{\bf e}_{ij}^\lambda (\hat n)\,h_{\lambda}(f, \hat n)\,,
\ee
imposing the condition
\be
\label{relcoa}
h_\lambda (-f, \hat n)\,=\,h^*_\lambda (f, \hat n)
\ee
  which ensures that $h_{ij}(t, \vec x)$ is real. We also assume that the polarization tensors ${\bf e}_{ij}^{\lambda}$ are real
quantities.  See Appendix \ref{AppA} for more details on our conventions. The presence of a
 GW deforms the geodesics of light, and produces a time delay $\Delta T_a(t)$ on the
period of a pulsar.
We denote with $\tau_a$ the time travelled
from a pulsar to the Earth, setting from now on $c=1$.  The pulsar is situated at the position 
(from now on, hat quantities correspond to unit vectors) 
\be
\label{defppa}
\vec x_a\,=\,\tau_a\,\hat x_a
\ee
 with respect
to the Earth, located at $\vec x\,=\,0$. The direction of the  vector $\hat x_a$
of the pulsar with respect to the Earth 
  plays an important role for our arguments.
 
 The time delay 
of the light geodesics 
reads
\bea
z_a(t)&=&\frac{\Delta T_a(t)}{T_a(t)}\,=\,\int_{-\infty}^{+\infty} d f\,e^{2 \pi i f t}\,z_a(f)
\\
&=&
\int_{-\infty}^{+\infty} d f\,e^{2 \pi i f t}\,\left(\sum_\lambda\,\int d^2 \hat n\,
D_a^{ij}(\hat n)\,{\bf e}_{ij}^\lambda (\hat n)\,h_{\lambda}(f, \hat n)
\right)\,.
\label{deftd}
\eea
 $D^{ij}_a$  is  the so-called detector tensor, controlling the connection
between  the light  delay~\footnote{
In writing eq \eqref{deftd}, we neglect as usual the `pulsar terms' and focus on the `Earth terms'.  We refer the reader to Chapter 23 of \cite{Maggiore:2018sht} and to the recent \cite{Romano:2023zhb} for  textbook discussions
on the quantities we are introducing, their physical properties, and more general information on how  PTA respond to GW physics.}
 and the GW:
\be
D^{ij}_a\equiv \frac{\hat x_a^i\,\hat x_a^j}{2(1+\hat n \cdot \hat x_a)}\,.
\ee
 Starting from the time delay $z_a(t)$, it is convenient to compute the time residual
\be
\label{deftra}
R_a(t)\,\equiv\,\int_0^t\,d t' z_a(t')\,,
\ee
which is easier to handle when Fourier transforming the signal. 
We assume that the correlators of Fourier modes of GW fluctuations can be expressed as
\be
\label{corrh1}
\langle
h_{\lambda}(f, \hat n)\,h^*_{\lambda'}(f', \hat n')
 \rangle 
 \,=\,\frac12\,S_{\lambda \lambda'}(f,\hat n)\,\delta(f-f')\,\frac{\delta^{(2)} (\hat n-\hat n')}{4\pi}
\,,
\ee
where 
the tensor $S_{\lambda \lambda'}(f,\hat n)$ defines the properties of the SGWB,
and $\lambda$ is the polarization index (we adopt a $(+,\times)$ basis
for the polarization tensors, see Appendix \ref{AppA}).  The quantity  $S_{\lambda \lambda'}(f,\hat n)$  can be decomposed in intensity and circular polarization as 
\be
\label{decsa}
S_{\lambda \lambda'}(f,\hat n)\,=\,
I(f,\hat n) \delta_{\lambda \lambda'}-i V(f, \hat n)\,\epsilon_{\lambda \lambda'}
\,,
\ee
where the $2\times2$ tensor $\epsilon_{\lambda \lambda'}$ is defined as  $\epsilon_{+ \times}\,=\,1\,=\,-\epsilon_{\times+}$, while $\epsilon_{+ +}\,=\,0\,=\,\epsilon_{\times \times}$. 
The SGWB intensity $I(f,\hat n) $ is real and positive,  and the circular polarization $V(f,\hat n) $ is a real quantity. 
Both quantities can depend on the GW frequency and direction, and behave as scalars under boosts. Our aim is to compute
 the response of a PTA system  to the presence of a SGWB, whose spectrum is characterized by (possibly anisotropic) intensity $I$ and circular polarization $V$ parameters. 

\smallskip

In order to do so, we compute the correlation among the time residuals of a pair of pulsars, denoted respectively by the letters $a$ and $b$,
which is  induced by the presence of GW. The
 correlations among time-residuals is essential for detecting and characterizing the SGWB \cite{Hellings:1983fr}. 
We introduce the short-hand  notation 
\be
D_a^{\lambda}(\hat n)\,\equiv\,
D_a^{ij}(\hat n)\,\,{\bf e}_{ij}^\lambda (\hat n)\,,
\hskip1cm;\hskip1cm
\Delta t_{12}\,=\,t_1-t_2\,,
\ee
and  compute the two-point correlators of the the two pulsar time delays 
 (we sum over repeated polarization indexes): 
 \bea
\langle z_a(t_1) z_b(t_2)
\rangle
&=& \int_{-\infty}^{\infty} d f \,d f' \,\int d^2 \hat n\,d^2 \hat n' 
D_a^{\lambda}(\hat n)
D_b^{\lambda'}(\hat n')\,e^{2 \pi i \left(f t_1+f' t_2 \right)}
\langle 
h_{\lambda}(f, \hat n)\,h_{\lambda'}(f', \hat n')
\rangle\,.
\nonumber\\
\eea

Making appropriate changes of variable, $f\to-f$ and $f'\to-f'$,
we can use  relation \eqref{relcoa} to recast the previous expression in a convenient form, which  allows
us to  apply eq \eqref{corrh1}:
\bea
\langle z_a(t_1) z_b(t_2)
\rangle
&=&\frac12 \,\int d f \,d f' \,d^2 \hat n\,d^2 \hat n' 
D_a^{\lambda}(\hat n)
D_b^{\lambda'}(\hat n')
\nonumber
\\
&&
\left( 
e^{2 \pi i \left(f t_1-f' t_2 \right)}
\langle 
h_{\lambda}(f, \hat n)\,h^*_{\lambda'}(f', \hat n')
\rangle
+
e^{-2 \pi i \left(f t_1-f' t_2 \right)}
\langle 
h^*_{\lambda}(f, \hat n)\,h_{\lambda'}(f', \hat n')
\rangle
\right)\,,
\nonumber
\\
&=&
\frac14\,\int d f  \,d^2 \hat n
\,D_a^{\lambda}(\hat n) D_b^{\lambda'}(\hat n)
\left( 
e^{2 \pi i f  \Delta t_{12}}\,
S_{\lambda \lambda'}(f,\hat n)
+
e^{-2 \pi i f \Delta t_{12}}\,
S^*_{\lambda \lambda'}(f,\hat n)
\right)\,,
\nonumber
\\
&=&
\frac14\,\int d f  \,d^2 \hat n
\,D_a^{\lambda}(\hat n) D_b^{\lambda'}(\hat n)
\nonumber
\\
&&
\left[ \cos{(2 \pi  f  \Delta t_{12})}\,\left( S_{\lambda \lambda'}(f,\hat n)+
S^*_{\lambda \lambda'}(f,\hat n)\right)+ i \sin{(2 \pi  f \Delta t_{12})}\,\left( S_{\lambda \lambda'}(f,\hat n)-
S^*_{\lambda \lambda'}(f,\hat n)\right)\right]
\,,
\nonumber
\\
&=&\frac12 \int{ d f}  \,d^2 \hat n
\,D_a^{\lambda}(\hat n) D_b^{\lambda'}(\hat n)
\left[ \cos{(2 \pi  f \Delta t_{12})}\, I (f,\hat n) \delta_{\lambda \lambda'} 
+\sin{(2 \pi  f \Delta t_{12})}\, V(f,\hat n) \epsilon_{\lambda \lambda'} 
\right]\,.
\nonumber\\
\eea
The previous result can be plugged in eq \eqref{deftra} to compute the two-point function of time residuals:
%
\bea
\langle 
R_a(t_A)
R_b(t_B)
\rangle
&=&\frac12 \int_0^{t_A} \int_0^{t_B}\,d t_1 d t_2\,\int d f \,d^2 \hat n
\nonumber
\\
&&
D_a^{\lambda}(\hat n) D_b^{\lambda'}(\hat n)
\,
\Big[ \cos{\left(2 \pi  f \Delta t_{12} \right)}\, I (f,\hat n) \delta_{\lambda \lambda'} 
+\sin{\left(2 \pi  f \Delta t_{12} \right)}\, V(f,\hat n) \epsilon_{\lambda \lambda'} 
\Big]
\nonumber
\\
&=&
\int \frac{d f\, \sin{\left( \pi f t_A \right)}\sin{\left( \pi f t_B \right)}}{\pi\,f^2}
\nonumber\\
&&\times
\left[\bar I(f)\,
\Gamma^I_{ab}(f)\,\cos{\left(2 \pi  f \Delta t_{AB} \right)}\,
+\bar V(f)\,\Gamma^V_{ab}(f)\,\sin{\left(2 \pi  f \Delta t_{AB} \right)}
\right]\,,
\label{twopcordt}
\eea
where $\bar I(f)$  is the isotropic value of the intensity integrated over all directions, while $\bar V(f)$
is its analog for circular polarization. 

\smallskip

 The  response functions of a pulsar pair to the GW intensity
and circular polarization are obtained integrating over  $\hat n$, and
read
\bea
\label{gammai1}
\Gamma^I_{ab}(f)&=&\frac{1 }{{2 \pi\,\bar I(f)}}\int d^2 \hat n\left(
D_a^{\lambda}(\hat n) D_b^{\lambda'}(\hat n)\,\delta_{\lambda \lambda'} \right)\,
{I (f,\hat n)} 
\,,
\\
\label{gammav1}
\Gamma^V_{ab}(f)&=&\frac{1 }{2\pi\,{\bar V(f)}}\int d^2 \hat n\left(
D_a^{\lambda}(\hat n) D_b^{\lambda'}(\hat n)\,\epsilon_{\lambda \lambda'} \right)\,
{V (f,\hat n)} 
\,.
\eea
These quantities depend on the relative position of the pulsars in the sky (given
the dependence of the quantities $D_{a,b}^\lambda$ on the pulsar location) and on  the
properties of the SGWB. Hence,
the response of a PTA system to GW depends on the pulsar configuration.
 Notice that only correlators \eqref{twopcordt} evaluated  at different times $t_A$ and $t_B$ are sensitive
 to
  circular polarization. In actual measurements though, correlators are weighted
  by suitable filters to better extract the signal.  
   We discuss in Appendix \ref{AppB}  and section  \ref{sec_strate}  
  the relation between  correlators as above and measurable GW signals, using a
  a match-filtering technique.

Our task~\footnote{While in this section and the next we focus
on spin-2 GW modes, the same procedure
can be applied to study alternative gravity models in which gravitation is mediated by a mixture of spin-2 and spin-0 or spin-1 fields. We  examine this possibility in section \ref{sec_MG}.} is to compute  $\Gamma^I_{ab}(f)$ and $\Gamma^V_{ab}(f)$. 
The results depend also on the theory of gravity one considers. For the case GW are carried by spin-2 fields, as in General Relativity, we can
 make use
of eqs  \eqref{impeq1} and \eqref{impeq2}.   The quantities within parenthesis
\footnote{Formulas similar to \eqref{contra} have been   used in \cite{Ali-Haimoud:2020ozu,Ali-Haimoud:2020iyz} to characterize scenarios with anisotropic GW intensity.  See also the treatment in \cite{Anholm:2008wy}. As far as we are aware, our formulas  for circular polarization are instead new in this context.}
 in eqs \eqref{gammai1}
and \eqref{gammav1} result: 
\bea
D_a^{\lambda}(\hat n) D_b^{\lambda'}(\hat n)\,\delta_{\lambda \lambda'}&=&\frac{(\hat x_a \cdot \hat n)^2+(\hat x_b \cdot \hat n)^2+ (\hat x_a \cdot \hat n)^2  (\hat x_b \cdot \hat n)^2-1}{8(1+\hat x_a \cdot \hat n)(1+\hat x_b \cdot \hat n)}
\nonumber\\
&+&\frac{ (\hat x_a \cdot \hat x_b)^2-2 (\hat x_a \cdot \hat x_b)  (\hat x_a \cdot \hat n)  (\hat x_b \cdot \hat n) 
}{4(1+\hat x_a \cdot \hat n)(1+\hat x_b \cdot \hat n)}
\,,
\label{contra}
\eea
and
\bea
\label{contrb}
D_a^{\lambda}(\hat n) D_b^{\lambda'}(\hat n)\,\epsilon_{\lambda \lambda'}
&=&\frac{\left[ \hat x_a \cdot \hat x_b 
- (\hat x_a \cdot \hat n)  (\hat x_b \cdot \hat n) 
\right] \left[ \hat n \cdot ( \hat x_a \times \hat x_b ) \right] }{{4(1+\hat x_a \cdot \hat n)(1+\hat x_b \cdot \hat n)}}
\,.
\eea
where $\times$ denotes cross product among vectors.
Plugging these results in eqs \eqref{gammai1}, \eqref{gammav1}, we are left with  angular integrals to carry on, which
depend on $I(f,\hat n)$
and $V(f,\hat n)$. The 
computation gives quantities
depending on the GW   
 frequency. We can already notice that, being the quantity in eq \eqref{contrb} an odd function of the vector
 directions, the integral over all directions give zero, unless the circular polarization function $V(f,\hat n)$ depend
 explicitly on the direction $\hat n$. Hence we need an anisotropic signal 
 to measure circular polarization \cite{Kato:2015bye,Belgacem:2020nda}, completely analogously to what happens 
 for planar interferometers (see e.g. \cite{Smith:2016jqs}).  
 
 \smallskip
 
 Starting from these basic formulas,
   in the
next section we apply them to the specific case of 
 spin-2 GW (including circular polarization), and study 
anisotropies induced by the motion of our reference
frame with respect to the rest frame of the SGWB source. 

\newpage
\section{PTA response to kinematic anisotropies: the spin-2 case}
\label{secptar}

In the previous section we developed a general treatment to investigate the PTA response to anisotropic GW signals.  We now specialise  our attention to the case of  Doppler anisotropies, in theories as General
Relativity
 where  GW are characterized by   spin-2 polarizations. Nevertheless, we include  also possible effects of parity violation \cite{Jackiw:2003pm,Alexander:2004us,Lue:1998mq,Satoh:2007gn,Contaldi:2008yz,Anber:2009ua,Alexander:2009tp,Takahashi:2009wc,Anber:2012du,Bartolo:2016ami,Mylova:2019jrj,Ozsoy:2021onx,Fu:2023aab}. The geometrical dependence  of the PTA response to kinematic
anisotropies make them
 a   promising probe of the source 
 of GW and of the physics of gravitation.
  
\smallskip
In fact,
kinematic anisotropies are a {\it guaranteed property} of a SGWB of primordial origin: their features are fully calculable, being  determined by the properties of the isotropic 
part of the SGWB. 
If the SGWB has a cosmological origin related with early-universe physics, we can expect that the relative motion among frames induces an 
 effect analog to the large dipolar anisotropy measured in the CMB \cite{Smoot:1977bs,Kogut:1993ag,WMAP:2003ivt,Planck:2013kqc}, whose amplitude  is a factor $1.2 \times 10^{-3}$ times smaller 
 than  the isotropic background. For the case of  the CMB, the dipolar kinematic anisotropy is well larger in size than its intrinsic anisotropies. The same can occur for a cosmological SGWB   detectable with PTA. For this reason it is worth characterizing kinematic effects, and explore what information we can extract from their detection  with PTA experiments.   We   now show  how the pulsar response to Doppler anisotropies depends 
on the pulsar location  in the sky;
 on the frequency dependence  of the isotropic part of the SGWB; and on the presence of parity violating effects in the SGWB as considered in various early universe scenarios. 

 %
 \smallskip
 
 Under a boost chararacterized by velocity
 \be
 \vec  v\, =\,\beta \hat v
 \ee
  (with $\beta=v/c=v$ in our units with $c=1$) the GW frequency scales as (see \cite{Cusin:2022cbb} and references therein)
 \be
 f\to f'={\cal D}^{-1}\,f\,,
 \ee
 with
 \be
 \label{defD}
{\cal D}\,=\,\frac{\sqrt{1-\beta^2}}{1-\beta \, \hat n\cdot \hat v}\,.
\ee

  The SGWB intensity $I(f,\hat n)$ and circular polarization $V(f, \hat n)$ (see eq \eqref{decsa}) are proportional to the GW energy density $\Omega_{\rm GW}(f, \hat n)$ through relations as
  $I \propto \Omega_{\rm GW}/f^3$  and $V \propto \Omega_{\rm GW}/f^3$. We
  will
assume that these quantities are isotropic in the source rest frame. They develop kinematic anisotropies in a  frame (like ours)  moving  with velocity $\vec v$ with respect to the source.  
   Since an isotropic $\Omega_{\rm GW}(f)$ scales  under
  boosts as $\Omega_{\rm GW}(f)\,\to\,\left(f/f'\right)^4\,\Omega_{\rm GW}(f')\,=\,{\cal D}^4\,
  \Omega_{\rm GW}\left({\cal D}^{-1} f \right)$ (see \cite{Cusin:2022cbb}), we find
  %
%
%
\bea
\frac{I(f,\hat n)}{\bar I(f)}\,=\,
\frac{{\cal D}\,
I({\cal D}^{-1}\,f)}{\bar I(f)}&=&
\left[1 +\frac{\beta^2}{6}\left(\alpha_I+n_I^2+2 n_I \right)\right]
\nonumber
\\
&&
-\beta n_I\, \hat n\cdot \hat v
+\frac{\beta^2}{2}\left[(\hat n\cdot \hat v)^2-\frac13 \right]\left( 
\alpha_I+n_I^2- n_I 
\right)
\label{defiexpa}
\,,
\\
\frac{V(f,\hat n)}{\bar V(f)}\,=\,
\frac{{\cal D}\,
V({\cal D}^{-1}\,f)}{\bar V(f)}&=&\left[1 +\frac{\beta^2}{6}\left(\alpha_V+n_V^2+2 n_V \right)\right]
\nonumber
\\
&&
-\beta n_V\, \hat n\cdot \hat v
+\frac{\beta^2}{2}\left[(\hat n\cdot \hat v)^2-\frac13 \right]\left( 
\alpha_V+n_V^2- n_V
\right)\,,
\label{defvexpa}
\eea
where  we introduce  parameters $n$ and $\alpha$ controlling the tilt of the isotropic background:
\bea
\label{defai}
\frac{d \,\ln \bar I(f)}{d \,\ln f}&=&n_I(f)+1
\hskip0.8cm,\hskip0.5cm
\frac{f^2\, \bar I''(f)}{\bar I(f)}\,=\,\alpha_I(f)+n_I(f)+n^2_I(f)
\\
\frac{d \,\ln \bar V(f)}{d \,\ln f}&=&n_V(f)+1
\hskip0.8cm,\hskip0.5cm
\frac{f^2\, \bar V''(f)}{\bar V(f)}\,=\,\alpha_V(f)+n_V(f)+n^2_V(f)
\label{defai2}
\eea
and the isotropic bar quantities $\bar I(f)$, $\bar V(f)$ are defined after eq \eqref{twopcordt}.
In writing equations \eqref{defiexpa} and \eqref{defvexpa} we expand  the definition of ${\cal D}$ of eq \eqref{defD} up to second
order in the expansion parameter $\beta$, and we assemble the results in a way that makes manifest how kinematic effects give rise to dipolar and quadrupolar
anisotropies~\footnote{They also give contributions of order $\beta^2$ to the monopole, a small effect that we  neglect from now on.}, as controlled by the size of $\beta$.  As suggested by CMB results, we expect $\beta$ to be of order $10^{-3}$, making the detection of kinematic anisotropies a demanding (but certainly interesting!) challenge for PTA experiments. 



\smallskip

From now on, in this section we consider
  two pulsars $a,b$ are located at positions $\vec x_a\,=\,\tau_a\,\hat x_a$ and $\vec x_b\,=\,\tau_b\,\hat x_b$
with respect to the Earth located  at the origin. 
The corresponding pulsar response functions $\Gamma_{ab}^{I, V}$  are analytically calculated
plugging the expressions \eqref{defiexpa} and \eqref{defvexpa} into  eqs \eqref{gammai1} and \eqref{gammav1}, and performing the angular integrals~\footnote{To carry
on the integrals we found convenient to make use of complex integration methods and Cauchy theorem, as explained in
detail in \cite{Jenet:2014bea}.}. The
results are easier to handle introducing the combination
\be\label{defyab}
{y}_{ab}\,=\,\frac{1-\hat x_a \hat x_b}{2}\,=\,\frac{1-\cos{\zeta}}{2}\,,
\ee
which 
depends on the relative angle  
 $\hat x_a \cdot \hat x_b\,=\,\cos{\zeta}
$
  of
the pulsar positional vectors with respect to the Earth (see eq \eqref{defppa}).  We now
analytically investigate how the integrals
depend on the pulsar positions with respect to the
velocity vector $\hat v$. We find exact results with a transparent geometrical interpretation,
which turns useful in developing strategies of detection in section \ref{sec_strate}. 

\subsection{Pulsar response to GW intensity}
\label{sec_rin}

The response function \eqref{gammai1} to the GW intensity, expanded up to
order  $\beta^2$, results
\bea
\label{respia}
\Gamma^I_{ab}&=&\Gamma_{ab}^{\rm HD}
+{\beta\,n_I}  F_{ab}^{(1)}+{\beta^2}\left( 
\alpha_I+n_I^2- n_I 
\right)\,F^{(2)}_{ab}\,,
\eea
with
\bea
\label{defFHD}
\Gamma_{ab}^{\rm HD}
&=&\frac13-\frac{y_{ab}}{6}+y_{ab} \ln y_{ab} 
\\
F^{(1)}_{ab}&=& 
\left(\frac{1}{12}+\frac{ y_{ab}}{2}+\frac{ y_{ab} \ln y_{ab}}{2(1-y_{ab})} \right)
\, \left[\hat v\cdot \hat x_a+\hat v\cdot \hat x_b\right]\,,
\label{defF11}
\\
F^{(2)}_{ab}&=&
\left(\frac{3-13  y_{ab}}{20 ( y_{ab}-1)}+\frac{ y_{ab}^2 \ln  y_{ab}}{2(1- y_{ab})^2} \right)
\,\left[(\hat v\cdot \hat x_a)(\hat v\cdot \hat x_b) \right]
\nonumber\\
&+&
\left( \frac{1+2  y_{ab}-4  y_{ab}^2+ y_{ab}^3+3  y_{ab} \ln  y_{ab}
}{
12 (1- y_{ab})^2
}\right)\,\left[(\hat v\cdot  \hat x_a)^2+(\hat v\cdot  \hat x_b)^2 \right]
\,.
\label{defF12}
\eea

The response function \eqref{respia}  includes a first part 
$\Gamma_{ab}^{\rm HD}$  (independent from $\beta$) corresponding
to the classic Hellings Downs curve \cite{Hellings:1983fr} which we collect in eq \eqref{defFHD}. A second part (first determined in \cite{Anholm:2008wy}) is
weighted by the relative velocity  $\beta$, and controls the dipolar kinematic 
anisotropy. 
As anticipated above,
 it is proportional to the slope $n_I(f)$ of the SGWB profile, as defined
 in eq \eqref{defai}, making Doppler effects a probe of features of 
 the SGWB frequency spectrum. Considering its geometrical properties, 
 we notice that eq \eqref{defF11} is  symmetric under the interchange of pulsar positions $a\leftrightarrow b$, 
and it vanishes
if the pulsar directions are orthogonal to the direction of the relative velocity $\hat v$ 
among frames. In fact, differently from  the  Hellings Downs case, the pulsar response to the kinematic
dipole depends on the relative angle between pulsars, but also on the angles made
by the pulsar directions with the vector controlling the velocity among frames \cite{Anholm:2008wy,Mingarelli:2013dsa}.
We find that  only pulsars whose direction vectors $\hat x$ have components along $\hat v$ respond to kinematic
anisotropies relative to  the GW intensity $I$. 

 A third part of  equation  \eqref{respia}, proportional to $\beta^2$, controls
 the kinematic quadrupole, and is suppressed by a factor $\beta^2$ with respect to the isotropic background.  The 
  quadrupole can be enhanced if the frequency dependence of the spectrum has features, which lead to a large parameter $\alpha_{I}$ defined
in eq \eqref{defai} and entering in eq \eqref{respia} (more on this later). Also the quadrupolar contribution to the response to the GW intensity
vanishes when the pulsar directions are orthogonal to $\hat v$. However, it
depends on a different way on the angles among pulsars directions and $\hat v$:
compare eqs \eqref{defF11} with \eqref{defF12}.

\subsection{Pulsar response to GW circular polarization}


The pulsar response function for a  circularly polarized GW signal  reads 
\bea
\label{respva}
\Gamma^V_{ab}&=&{\beta\,n_V}\,G^{(1)}_{ab}
+{{\beta^2} \left( 
\alpha_V+n_V^2- n_V 
\right)}    
G^{(2)}_{ab}\,,
\eea
with
\bea
G^{(1)}_{ab}&=&-\left(\frac13+\frac{y_{ab} \ln y_{ab}}{4(1-y_{ab})} \right)\,\left[ \hat v\cdot (\hat x_a\times \hat x_b)\right]
\nonumber
\\
\label{respvaA}
G^{(2)}_{ab}&=&\left(
  \frac{ 2 y_{ab}^2-1-y_{ab} -3 y_{ab} \ln y_{ab}}{12 (1-y_{ab})^2}
  \right)
  \,\left[\left(\hat v\cdot  \hat x_a+\hat v\cdot  \hat x_b \right)\left(\hat v\cdot ( \hat x_a\times  \hat x_b)  \right)\right]
  \label{defG12}
\eea

\begin{figure}
\begin{center}
    \includegraphics[width = 0.47 \textwidth]{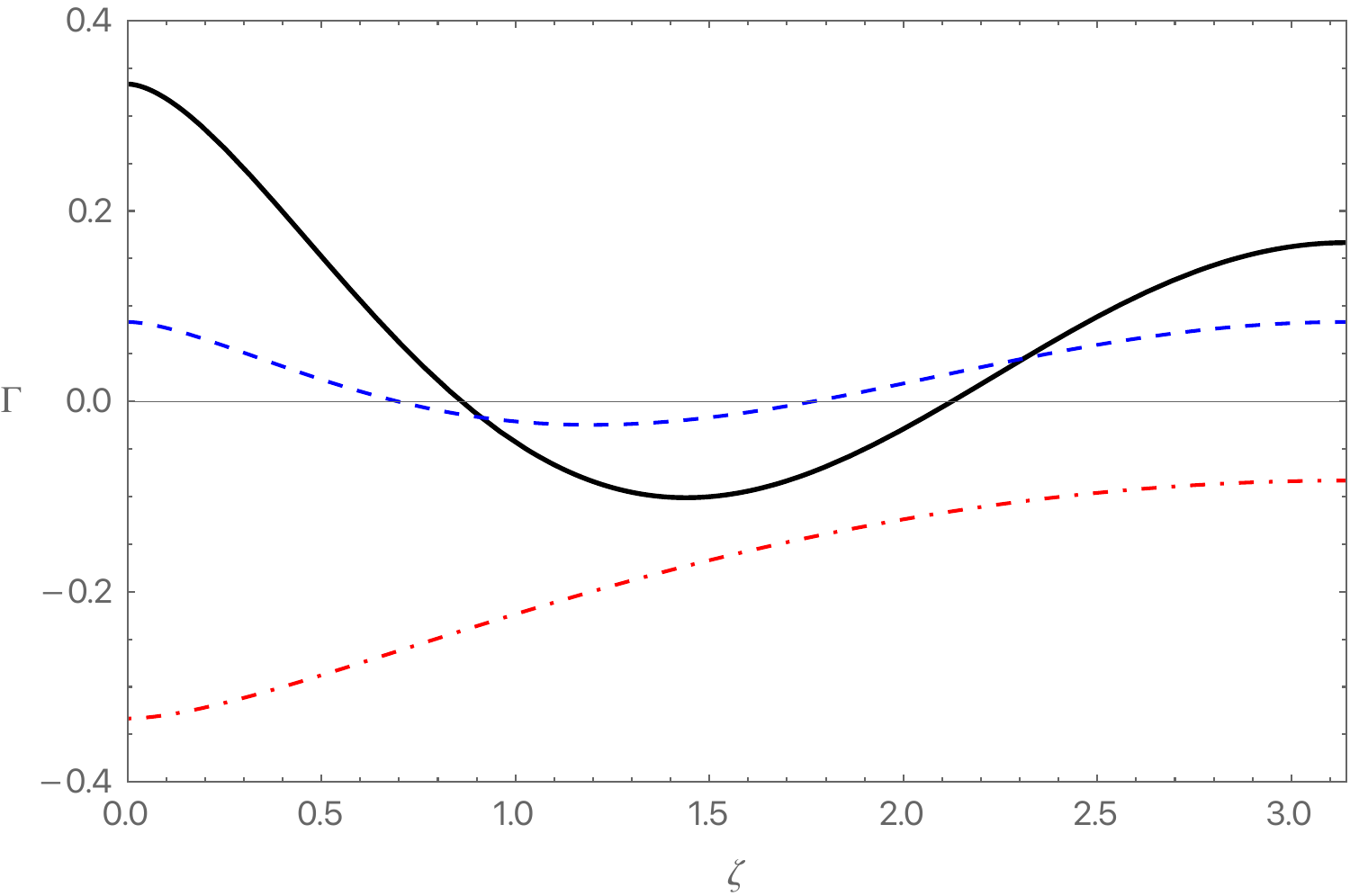}
  \includegraphics[width = 0.47 \textwidth]{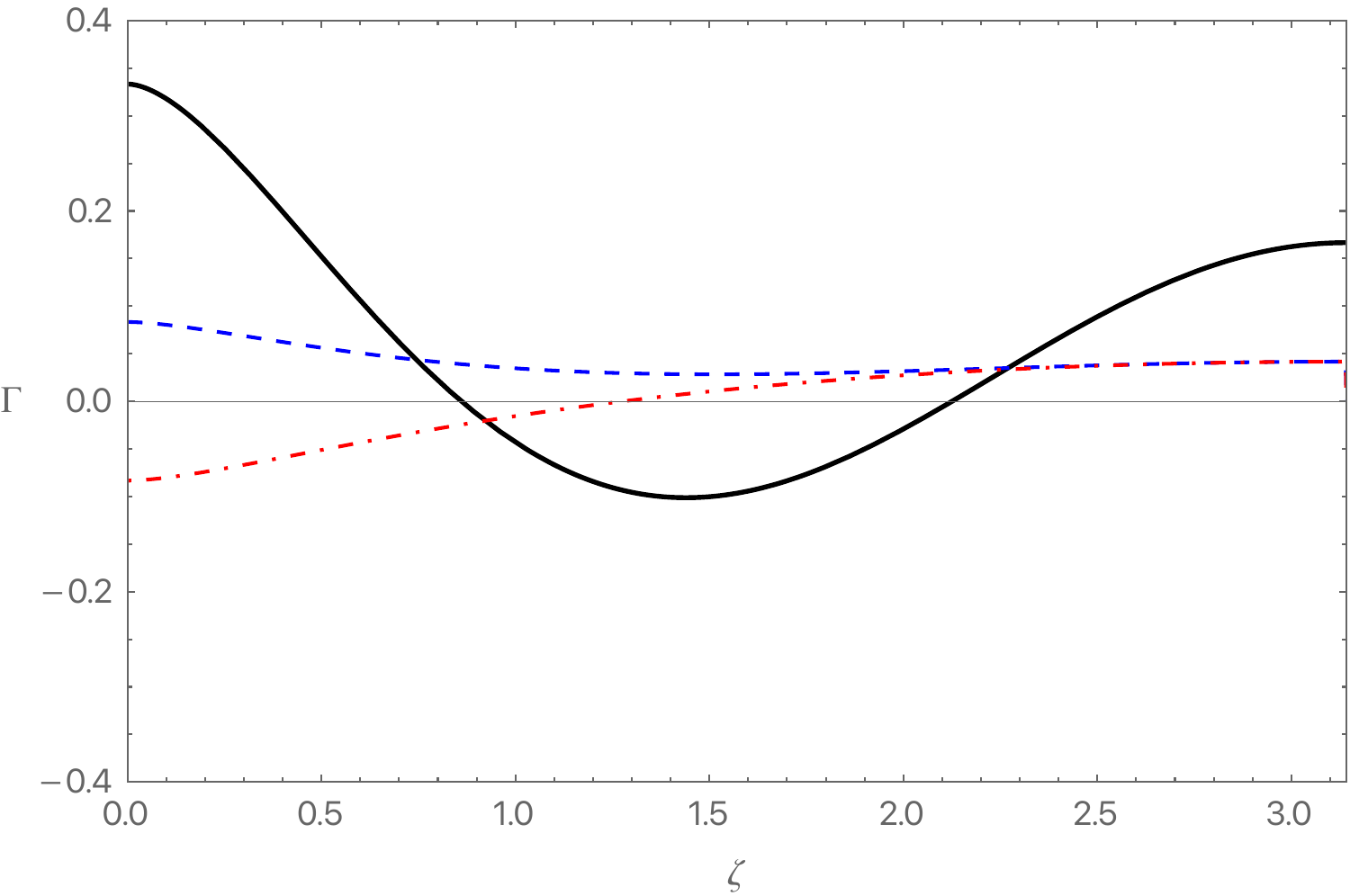}
    \end{center}
 \caption{ \small We represent  those contributions to PTA response functions
 to GW intensity and circular polarization, which depend on the angle $\zeta$ among pulsar directions. Namely, 
 we plot the quantities within round parenthesis of eqs \eqref{defF12} and
 \eqref{defG12}. In black, the Hellings-Downs curve.
 {\bf Left panel:} dashed blue:  dipole response to  GW intensity; dot-dashed red:  dipole 
  response to  GW circular polarization. {\bf Right panel:}
  dashed blue:  quadrupole response to  GW intensity (the part proportional to
  $(\hat v \cdot p_\alpha)^2+(\hat v \cdot p_\beta)^2$); dot-dashed red:  quadrupole 
  response to  GW circular polarization. 
 } 
  \label{fig_confinfAaa}
\end{figure}

\medskip

The response function \eqref{respva}   starts at order $\beta$: the PTA response 
to circular polarization vanishes for an isotropic background, and we need to include SGWB anisotropies for
being sensitive to this quantity \cite{Kato:2015bye}. The geometrical reason being an integration over all directions, which vanishes for parity-odd isotropic signals associated with circular polarization (see comment
after eq \eqref{contrb}).
 This is similar in spirit to what happens
for planar interferometers as LISA, for which we need to probe anisotropies 
in order to measure effects due to parity violation \cite{Seto:2008sr,Seto:2006hf,Seto:2006dz,Domcke:2019zls}.

Also in this case, kinematic anisotropies are proportional to parameters
controlling the frequency slope of the circular polarization function. 
The expression \eqref{respva} demonstrates that
the PTA pair can detect circular polarization only if $\hat v$ has components {\it orthogonal} to the plane formed by the pulsar directions. This feature is opposite to what found above when discussing pulsar response to GW intensity. This property -- which we point out for the first time --  can then be useful for experimentally
distinguish the two contributions: see section \ref{sec_strate}.

\smallskip

It is not straightforward to represent in a plot the rich, multi-parameter dependence
of the response functions on the geometrical configuration of the system. For definiteness, 
in Fig \ref{fig_confinfAaa}, left panel, we represent the pulsar angular response to the kinematic dipole, both for intensity and circular polarization, as a function
of the angle between pulsar directions.  In the right panel, we
represent the response functions to the kinematic quadrupole.
We plot only
 the 
quantities between round parenthesis in eqs   \eqref{defF12} and
 \eqref{defG12}, without the geometrical
factors depending on the relative velocity $\hat v$ among frames. A more detailed analysis of the geometrical 
dependence on the various quantities is postponed to section \ref{sec_strate}. For the moment, it is sufficient to notice that the dependence on the pulsar separation
of the functions $F^{(i)}_{ab}$ and $G^{(i)}_{ab}$ is qualitatively different from the Hellings-Downs curve,
which we also represent for comparison.

\subsection{Examples of SGWB frequency dependence}
\label{sec_someex}

Besides the geometrical dependence of the results on the location of
pulsars, the PTA response to Doppler effects depends on the
frequency dependence of the isotropic part of the SGWB.
 In fact,
the size and observational impact of kinematic anisotropies  can be enhanced
if the SGWB is characterised by large values for the slope parameters $n$ and
$\alpha$ of eqs \eqref{defai}, at least for certain ranges of frequencies. While current 
 PTA data are compatible with a power-law profile of the isotropic SGWB as
 function of frequency,  in the future 
 more
 accurate data might favour other scenarios. For example, 
   broken power law profiles, well 
  motivated by a variety of  SGWB sources \cite{Kuroyanagi:2018csn}, have also been considered in \cite{NANOGrav:2023gor} for
   explaining the most recent pulsar timing array  data (see also \cite{Ye:2023xyr} for more complex profiles in frequency). 
   The same is true  for 
     more complex  SGWB profiles motivated by primordial black hole physics: see e.g. the first analysis carried out in  \cite{NANOGrav:2023gor,NANOGrav:2023hvm,EPTA:2023xxk,Franciolini:2023pbf,Figueroa:2023zhu}  (and also
\cite{Ozsoy:2023ryl}    for a  recent review on inflationary primordial black hole models). 

We briefly outline the consequences of  these two scenarios for the physics of   kinematic anisotropies:

\begin{figure}
\begin{center}
    \includegraphics[width = 0.49 \textwidth]{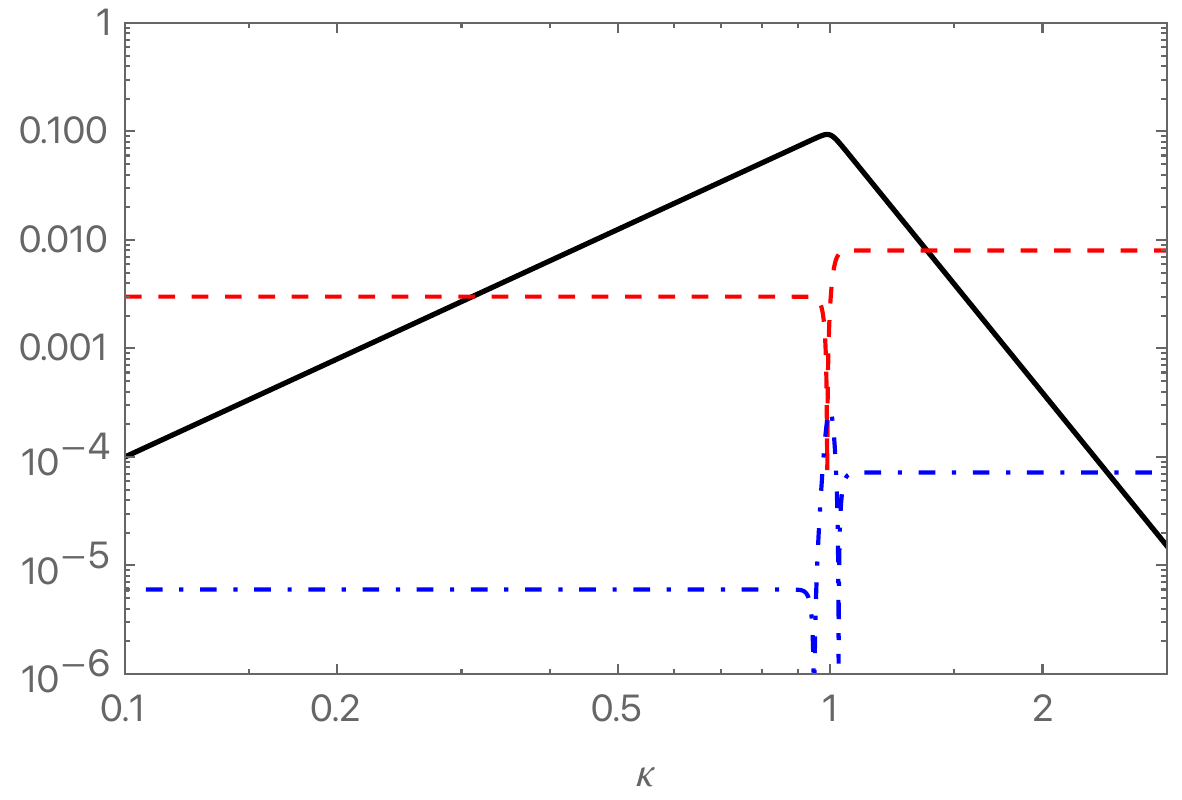}
        \includegraphics[width = 0.49 \textwidth]{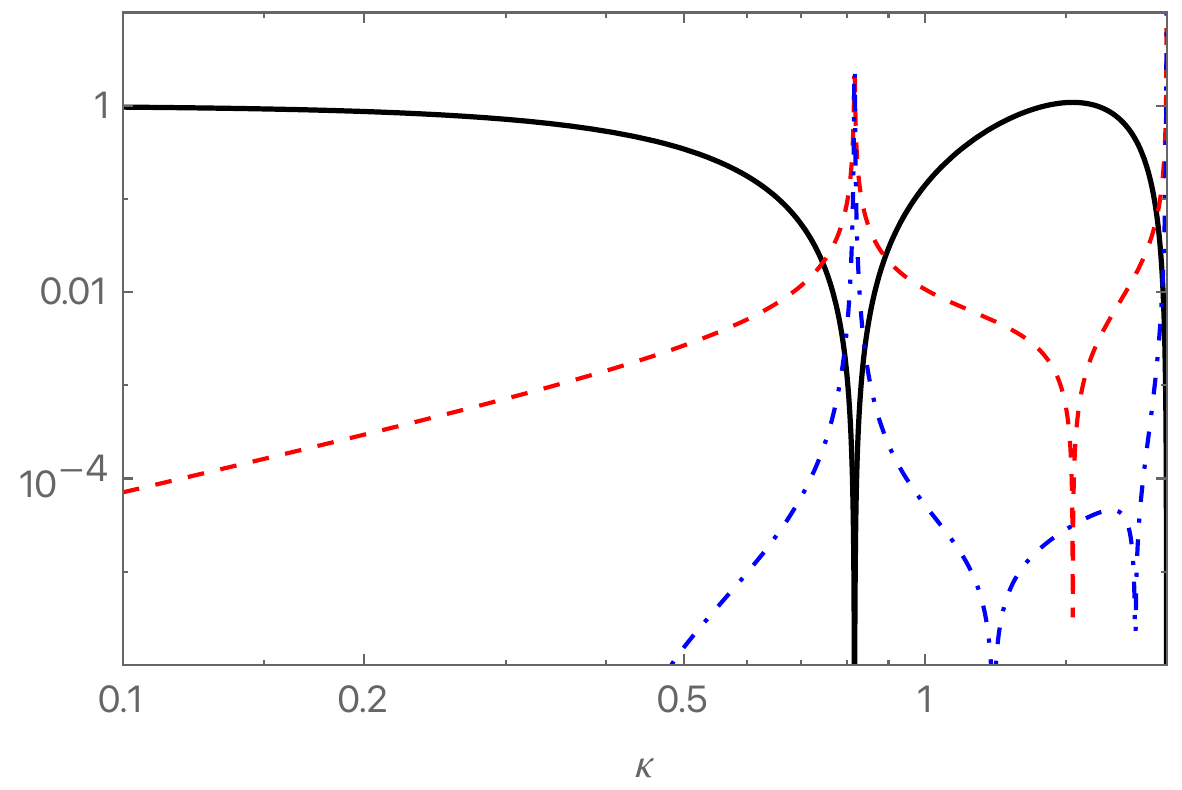}
    \end{center}
 \caption{ \small  Examples of intensity spectrum profiles, and combinations
 of slope parameters characterizing kinematic anisotropies. In all cases, $\beta=1.2 \times 10^{-3}$.
 {\bf Left panel}: broken power law profile of eq \eqref{ansint1}. 
 We choose $ I_0 = 0.1$,   $\sigma = 100$,  $\gamma = 3$, and  $\delta = 8 $. Black line: the intensity. Dashed red line: the combination $\beta |n_I|$ characterizing the amplitude of the dipolar anisotropy. Dot-dashed blue line: the combination  $\beta^2 |\alpha_I+n_I^2-n_I|$, characterizing the amplitude of the quadrupolar anisotropy. Notice that in proximity of the break of the power law the amplitude of the quadrupolar anisotropy is comparable to the dipolar one. 
  {\bf Right panel}: Ansatz \eqref{ansPBH} motivated by the physics of primordial 
  black holes. Color codes as above. In proximity of dips of intensity, the size of kinematic anisotropies
  become large. 
 }
  \label{fig_confinfaab21a}
\end{figure}

\smallskip
\noindent
{\bf Broken power law:} 
   Consider the following broken power law Ansatz for the  intensity profile
  \cite{Sampson:2015ada,Kaiser:2022cma}
   \bea
\label{ansint1}
\bar I(f)\,=\,I_0\,\kappa^{\gamma+1} \,\left[1+\kappa^\sigma\right]^{-\frac{\gamma+\delta}{\sigma}}
\,,
\eea
where we denote $\kappa=f/f_\star$,
and $I_0$, $\gamma$, $\delta$, $\sigma$ are positive dimensionless parameters, while $f_\star$ a pivot frequency  within the PTA detection band. The associated tilt parameters result
\bea
n_I&=&\frac{\gamma-\delta\,\kappa^\sigma}{1+ \kappa^\sigma}
\,,
\nonumber
\\
\alpha_I&=&-\frac{\sigma(\gamma+\delta)\, \kappa^\sigma}{\left( 1+ \kappa^\sigma\right)^2}
\,.
\label{anstil2}
\eea
The resulting slope $n_I$ of the intensity spectrum is equal to  the parameter $\gamma$ at small frequencies, and to $-\delta$ at larger frequencies. For a large
values of $\sigma$, at intermediate frequencies the quantity $\alpha_I$ can increase by several orders of magnitude, enhancing
the quadrupolar kinematic anisotropies to a size comparable to the dipolar ones (see \cite{Dimastrogiovanni:2022eir} for related effects explored for the case of interferometers).  We represent
in Fig \ref{fig_confinfaab21a}, left panel, an example to visually demonstrate such behaviour for
the slope parameters.

\smallskip
\noindent
{\bf Second-order GW from primordial black hole formation:} Primordial black
holes can form in the early universe.   Their formation requires to amplify the spectrum 
of scalar fluctuations for a range of    scales around a pivot frequency $f_\star$, which we
can consider in the PTA frequency ranges. At second 
order in perturbations, these mechanisms induce a SGWB, whose properties
depend on the source scalar spectrum \cite{Ananda:2006af,Baumann:2007zm,Saito:2008jc,Saito:2009jt} (see also
\cite{Domenech:2019quo,Domenech:2020kqm,Domenech:2021ztg}).  For a  narrow scalar spectrum whose width is
much smaller than $f_\star$, we expect that the GW energy density $\Omega_{\rm GW}(f)$
increases as $(f/f_\star)^3$ for small frequencies $f\ll f_\star$ \cite{Cai:2019cdl}. Then its profile has a dip (typically at scales around 
$f/f_\star \simeq \sqrt{2/3}$), followed by a pronounced resonance. The frequency profile  then drops its amplitude for  $f>2f_\star$ (see the discussion in \cite{Saito:2008jc,Saito:2009jt}). Given that intensity is connected to the GW energy density through the relation $f^3\,\bar I(f)\propto\Omega_{\rm GW}(f)$, we can consider the following simple Ansatz for the intensity parameter (we call again
 $\kappa=f/f_\star$) 
\be\label{ansPBH}
I(\kappa)\,=\,\frac{(\kappa^2-4)^2\,(3 \kappa^2-2)^2}{64}\hskip0.8cm,\hskip0.8cm 0\le\kappa\le2
\,,
\ee
as representative of the behaviour described above. In fact, the intensity $I(\kappa)\propto \Omega_{\rm GW}(\kappa)/\kappa^3$ is constant at small frequencies $\kappa\ll1$, for then
developing the features described above.  
Hence, in this context,
\bea
n_I&=&\frac{8\, \kappa^2 \,(3 \kappa^2-7)}{3 \kappa^4-14 \kappa^2+8}
\,,
\\
\alpha_I&=&-\frac{16\, \kappa^4\, (9 \kappa^4-42 \kappa^2+74)}{\left(3 \kappa^4-14 \kappa^2+8 \right)^2}\,.
\eea
We represent in Fig \ref{fig_confinfaab21a}, right panel, the resulting intensity and the slope parameter
combinations characterizing kinematic anisotropies. In proximity of the dip in intensity which preceeds the resonance, the size of the  slope parameters become so large that dipolar and quadrupolar anisotropies
become comparable in size.

\section{Kinematic anisotropies and modified gravity}
\label{sec_MG}

Theories of  gravity alternative to General Relativity allow  for scalar  and vector
metric components  to participate to gravitational interactions, and to contribute
to the GW signal in terms of extra GW polarizations \cite{Eardley:1973br}.  The PTA responses to scalar and vector polarizations
have been derived  in \cite{Lee_2008,Gair:2012nm,Chamberlin:2011ev,Shao:2014wja,Gair:2015hra,Hotinli:2019tpc,Liu:2022skj} for the case of  isotropic backgrounds (see also the general treatment in \cite{Romano:2016dpx}). Such
  extra polarizations induce correlations among
pulsar signals distinct from the Hellings Downs curve. These alternative predictions  
are not favoured by recent detections \cite{NANOGrav:2023gor}, that can be used
to set upper bounds on scalar and vector contributions to the isotropic backgrounds. Nevertheless, in
view of  future opportunities to further improve constraints (or maybe even detect  signals of modified gravity), it is interesting to  develop
this topic, and analyse how scalar and vector GW polarizations contribute
to kinematic anisotropies.

 Analogously to the spin-2 case, the PTA response to kinematic
 anisotropies associated with extra GW polarisations are sensitive to the slope  of scalar and vector spectra,
 as well as to the presence of parity violating effects in the vector sector. 
 Doppler effects are  unique probes of these features, which can increase the opportunities
 of detection. Here we derive for the first time analytical formulas describing the PTA
 response functions to kinematic anisotropies, associated with scalar and vector polarizations. 

\smallskip

The procedure to follow for determining the PTA response to kinematic anisotropies is identical to what done in sections \ref{secsetup} and \ref{secptar} for the tensor case -- the
only difference being that we utilize  the corresponding scalar and vector GW
polarization  tensors, whose properties are described in Appendix \ref{AppA}. 
 In particular, the scalar and vector polarization tensors are used to perform angular integrals corresponding to eqs \eqref{gammai1} and \eqref{gammav1}.

\begin{figure}
\begin{center}
    \includegraphics[width = 0.47 \textwidth]{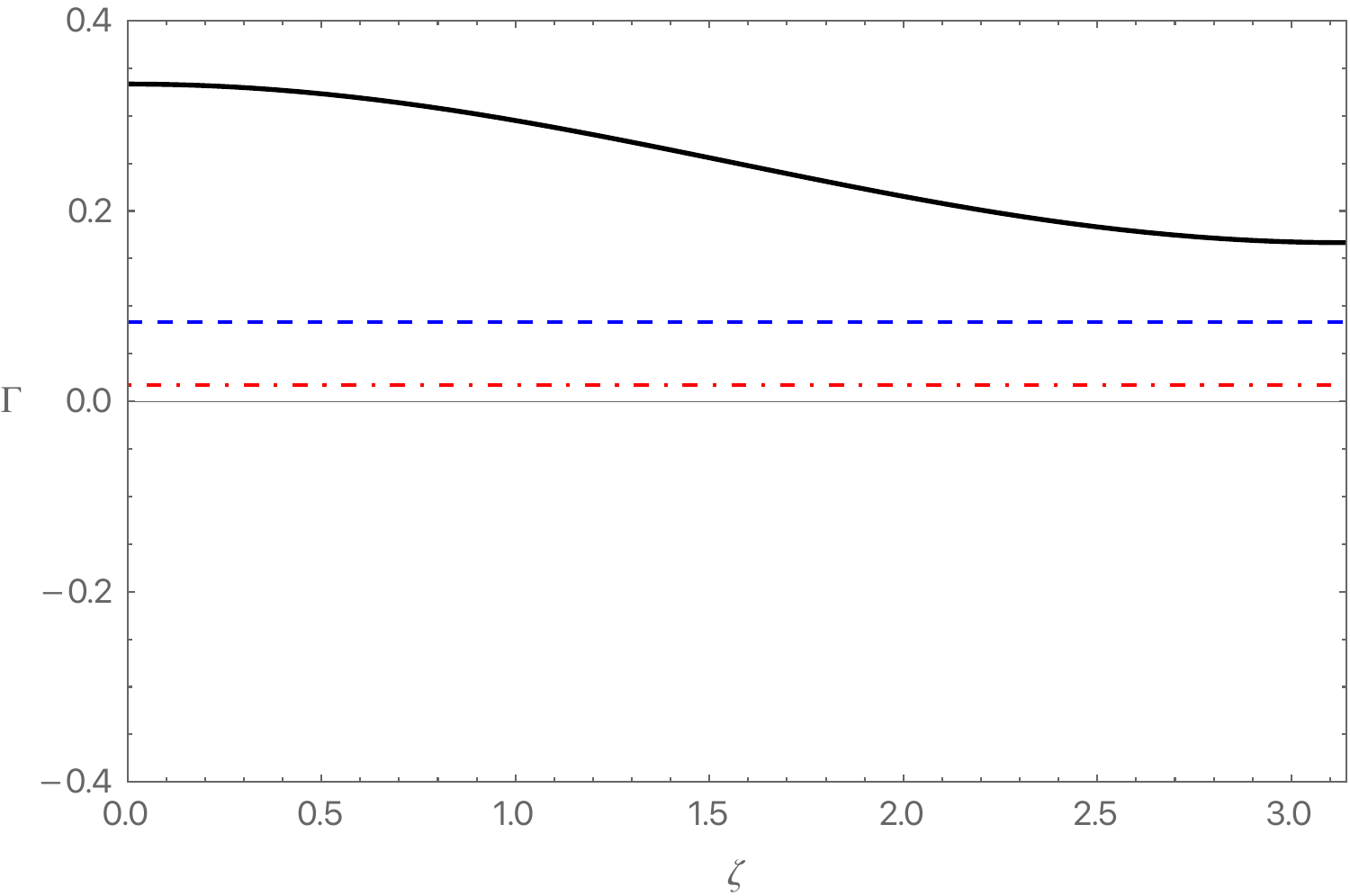}
  \includegraphics[width = 0.47 \textwidth]{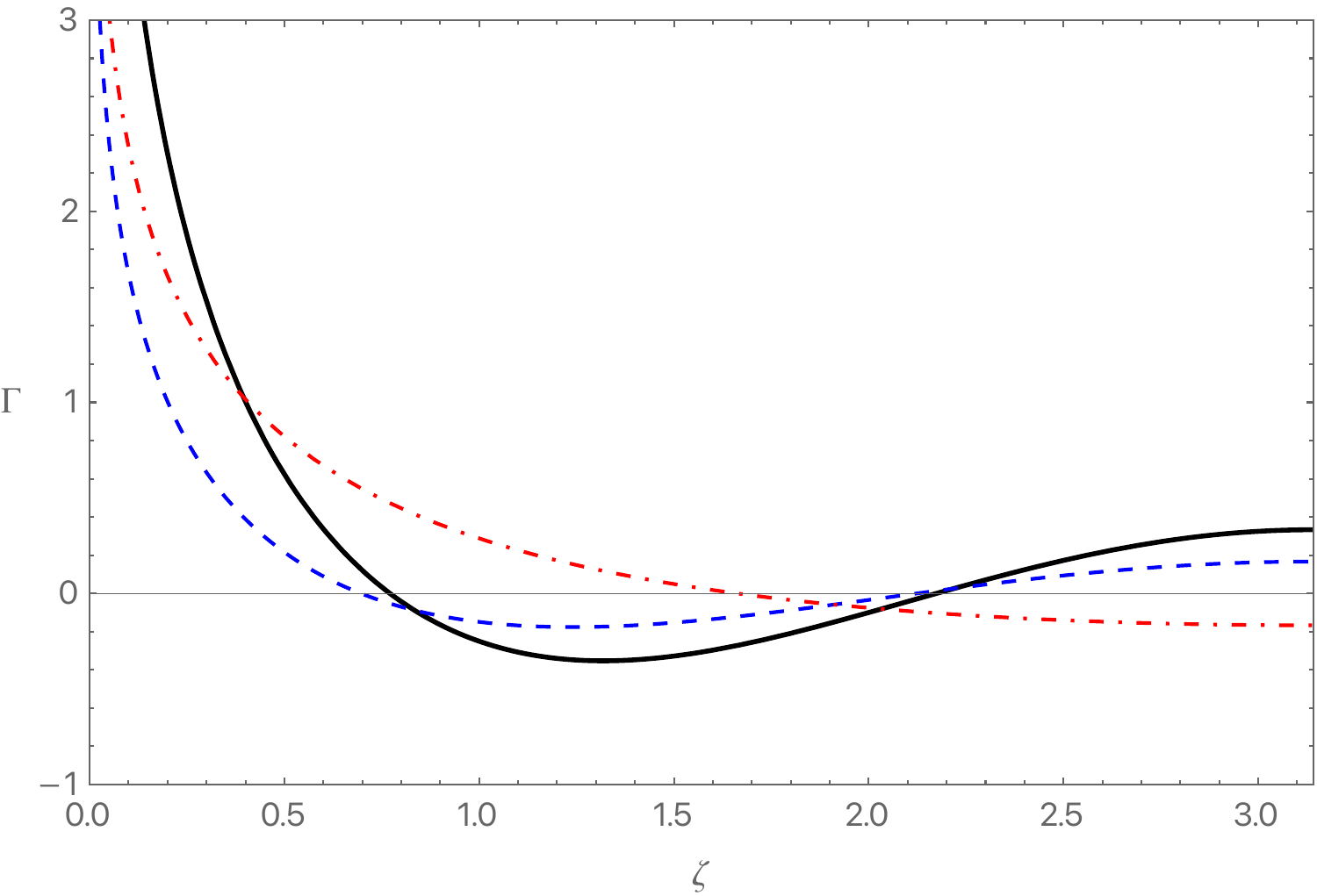}
    \end{center}
 \caption{ 
 \small We represent  part of the PTA response function
 to GW intensity and circular polarization in modified gravity, including kinematic anisotropies,
 as a function of the angle $\zeta$ among pulsar directions.
 We understand
the overall factors depending on the slopes $n$ and $\alpha$,
as well as the geometrical factors depending on the velocity among frames. 
  {\bf Left panel:} Black line: PTA response to an isotropic scalar GW component.
  Dashed blue:  dipole response to  scalar GW intensity. Dot-dashed red:  quadrupole 
  response to scalar  GW intensity. {\bf Right panel:} Black line: PTA response to an isotropic vector  GW component.
  Dashed blue:  dipole  response to  vector GW intensity.  Dot-dashed red:  dipole 
  response to  vector GW circular polarization. 
  In the right plot,
 we plot the quantities within round parenthesis of eq \eqref{GammaVe1}.
 } 
  \label{fig_confinfAaaa}
\end{figure}

\smallskip 
Scalar contributions to GW signal are characterized only by scalar intensity, and there
is no circular polarization.   We can
assume that the total isotropic part for the  intensity $\bar I(f)$ of the SGWB  is made of a tensor 
and a scalar part, 
$\bar I(f)\,=\,\bar I_{\rm tn}(f)+\bar I_{\rm sc}(f)$, 
possibly hierarchical in size as $\bar I_{\rm sc}\ll \bar I_{\rm ten}$. We
 use the same notation of section \ref{secptar}. We introduce the slope parameters
 for each of the intensity contributions
\bea
\label{defnsc}
\frac{f\,d  \bar I_{\rm tn}(f)}{\bar I_{\rm }(f)\,d  f} &=&n_{tn}+1\hskip0.7cm,\hskip0.7cm 
\frac{f\,d  \bar I_{\rm sc}(f)}{\bar I_{\rm sc}(f)\,d  f} \,=\,n_{sc}+1 \,,
\\
\label{defasc}
\frac{f\,d  \bar I^2_{\rm tn}(f)}{\bar I_{\rm }(f)\,d  f^2} &=&
\alpha_{tn}+n_{tn}^2+n_{tn}\hskip0.7cm,\hskip0.7cm 
\frac{f\,d  \bar I^2_{\rm sc}(f)}{\bar I_{\rm }(f)\,d  f^2} \,=\,
\alpha_{sc}+n_{sc}^2+n_{sc} \,.
\eea
  We find that
the PTA response function to scalar part of the intensity is
\be
\label{GammaSc1}
\Gamma_{ab}^{\rm sc}\,=\,\frac13-\frac{y_{ab}}{6}+\beta\,\frac{n_{\rm sc}}{12}\,
\left[ 
\hat v\cdot  \hat x_a+\hat v\cdot  \hat x_b
\right]
+
\frac{\beta^2}{60}\left( 
\alpha_{\rm sc}+n_{\rm sc}^2- n_{\rm sc}
\right)\,\,\left[(\hat v\cdot \hat x_a)(\hat v\cdot \hat x_b) \right]\,,
\ee
where, beside the isotropic part, we  consider the dipolar (proportional to $\beta$)
and quadrupolar (proportional to $\beta^2$) kinematic anisotropy contributions. (Instead,
the tensor part is identical to what discussed in section \ref{secptar}.)
 Interestingly --  and differently   from  the tensor case -- the PTA response to scalar
kinematic anisotropies is independent from the angle $\hat x_a \cdot \hat x_b$ among
pulsars, while it depends on the angles among the pulsar directions and the frame velocity 
vector $\hat v$.  See Fig \ref{fig_confinfAaaa}, left panel.  Moreover, it is proportional
to $n_{\rm sc}$, making it a unique probe of the slope of the scalar contribution to the
total intensity (more
on this in section \ref{sec_strate}).

\smallskip

Vector-tensor non-minimally coupled theories of gravity have also been introduced in view of
applications to dark energy and black hole physics (see e.g. \cite{Tasinato:2014eka,Heisenberg:2014rta,Tasinato:2014mia,Chagoya:2016aar}). The correlators
among vector Fourier modes can be decomposed into intensity and circular 
polarization, analogously to what done in the tensor case in section  \ref{secsetup}. Using
formulas \eqref{impeq1v} and \eqref{impeq2v}, it is straightforward to compute the PTA response 
to  dipolar anisotropies of the vector intensity and vector circular polarization, finding (the slope parameters are defined analogously as above)
\bea
\Gamma_{ab}^{\rm vector}&=&
-\frac73+\frac{8 y_{ab}}{3}-\ln y_{ab}
\nonumber
\\
&
+&
{n_{\rm vec}\,\beta} \left\{ \left(y_{ab}-\frac43-\frac{ \ln y_{ab}}{2 (1-y_{ab})}
\right) \left[ 
\hat v\cdot  \hat x_a+\hat v\cdot  \hat x_b
\right]+ \left(-\frac23-\frac{ \ln y_{ab}}{2 (1-y_{ab})}
\right)\left[ \hat v \cdot (\hat x_a\times \hat x_b)\right]\right\}\,.
\nonumber
\\
\label{GammaVe1}
\eea

The response to vector GW components does depend on the angular separation among pulsar directions.  See Fig \ref{fig_confinfAaaa}, right panel. It is very large for small angular separation
among pulsars, where contributions of pulsar terms should nevertheless
be included, see e.g. \cite{Lee_2008,Gair:2014rwa}. In the second line of eq \eqref{GammaVe1} we find  in the first
the pulsar dipolar response to vector intensity, that depends on the projection of the pulsar
directions on the velocity vector $\hat v$. The second term controls the the pulsar dipolar response to vector circular polarization, depending on the projection of $\hat v$ on  a direction perpendicular to the plane of the two pulsar directions. These features
are  similar to what found for the spin-2 case, see  section \ref{secptar}.

\section{Strategies of detection}
\label{sec_strate}

So far, we learned several interesting features of kinematic anisotropies,
which make them potential probes of specific properties of the SGWB:
\begin{itemize}
\item \label{pointone} The geometrical configuration of the pulsar network 
enters in the PTA response functions, which depend
in a distinctive way on intensity and circular polarization of 
the SGWB. While the response to GW intensity is maximal for pulsars
whose  positional vectors are in the direction of  the Doppler velocity vector $\hat v$,
the response to GW circular polarization is enhanced for pulsars lying on a plane orthogonal 
to  $\hat v$. This distinct   behaviour in
the two cases can be used
 for selecting  PTA sets aiming at distinguishing and independently measure
the  GW intensity and circular polarization. 
\item The overall size of kinematic anisotropies can be amplified in scenarios with a rich   frequency dependence of the isotropic part of the background, see e.g.
section \ref{sec_someex}. This properties applies to GW backgrounds associated with standard spin-2 polarizations, but possibly also containing  GW spin-0 and spin-1 contributions.  Hence,
a detection of kinematic anisotropies can give us a complementary probe of the slope
of  GW polarization components, which is  difficult  or even impossible to extract  from 
the study of the isotropic part of the background only.
\end{itemize}
We now make  concrete applications 
of our findings,  in order to develop  specific strategies of detection of different physical effects.
In a sentence,
the main feature we can exploit is the following:
\begin{itemize}
\item[$\triangleright$] 
The presence of
 kinematic anisotropies implies
 that signal correlations among pulsars do not follow the
standard Hellings Downs angular distribution, but have deviations
that depend on SGWB  properties. Measurements of these
deviations, that are {\bf deterministic} and  {\bf analytically calculable},  can be used to extract information
about the GW signal.
\end{itemize}

 As explained in the Introduction, we are especially interested in using kinematic anisotropies as a probe of: 1.
 parity violation,
 2. 
  the frequency slope
 of correlation functions, 
 3. 
 the presence of extra polarizations. 
We discuss these three topics in what follows.

\subsection{Kinematic anisotropies and SGWB  circular polarization}
\label{sec_str_cp}

The effects of  of parity violation in the SGWB kinematic anisotropies can be cleanly extracted from PTA data, once we
make an appropriate choice of pulsars to monitor. In fact, it is not even necessary to make combinations among
different signals, as  proposed in the first work \cite{Kato:2015bye} discussing the topic (see also \cite{Belgacem:2020nda}). Choosing carefully the pulsar system, we can 
reduce sources of statistical errors by focussing only on pulsars located at convenient positions to reveal the presence of parity violation. 
Eqs
\eqref{respva} and \eqref{respvaA} teach us
that  PTA are sensitive to circular polarization only through the anisotropies
of the SGWB
 (as first pointed out in \cite{Kato:2015bye}), and moreover the corresponding
 signal is enhanced when maximising the quantity $\left[ \hat v\cdot (\hat x_a\times \hat x_b)\right]$. 
(The same is true for parity violating effects in possible spin-1 polarization contributions, see section \ref{sec_MG}.) 

\begin{figure}
\begin{center}
        \includegraphics[width = 0.47 \textwidth]{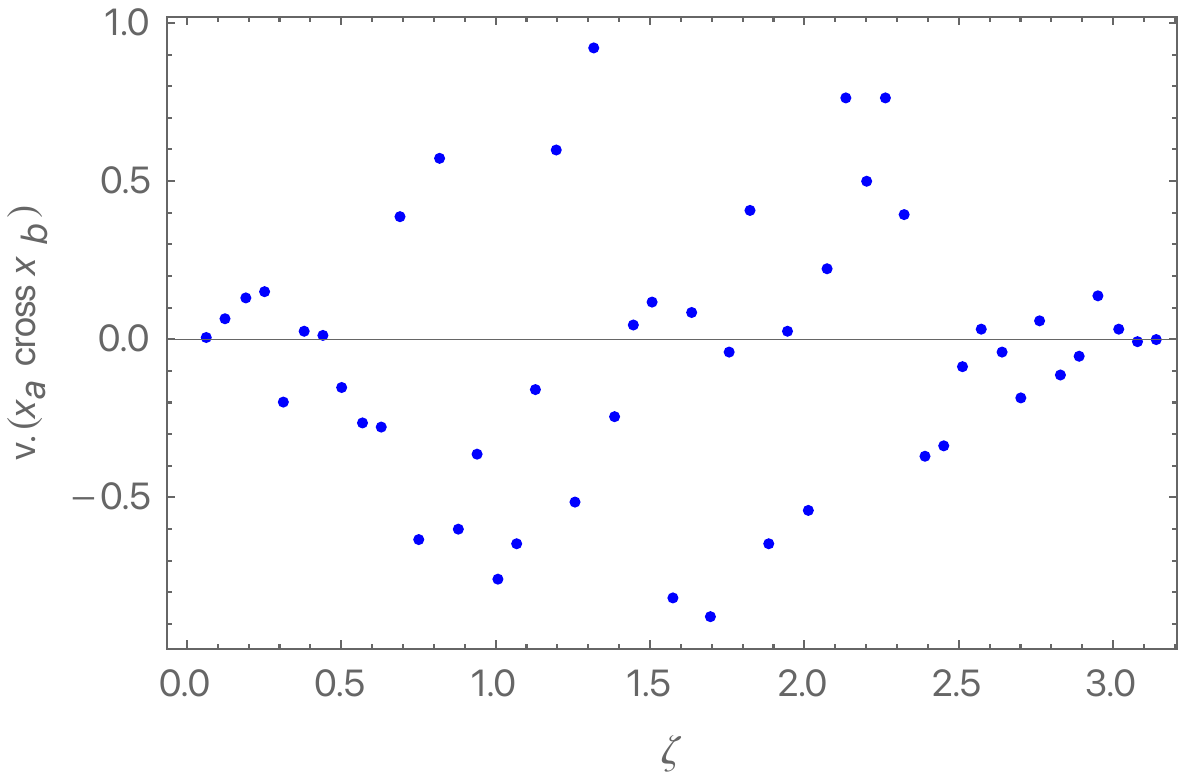}
    \includegraphics[width = 0.47 \textwidth]{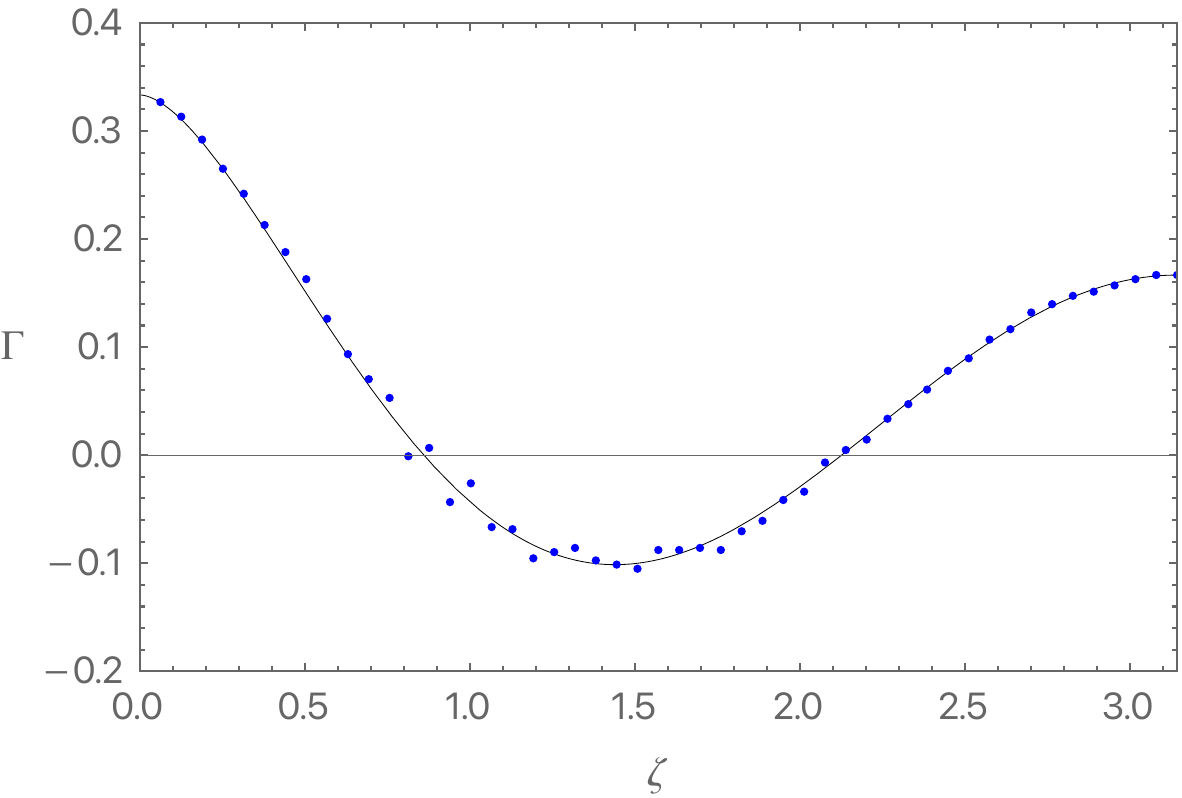}
    \end{center}
 \caption{ 
 We consider a random collection of pulsar pairs whose positional
 vectors $\hat x_a$ and $\hat x_b$ lie on the a plane orthogonal to the relative velocity $\hat v$. Our aim is
 to represent how geometrical factors depending on $\hat v$ affect the response functions for circular polarization, as a function of the angle $\zeta$ among pulsars.
 {\bf Left } combination 
 $\hat v\cdot (\hat x_a\times \hat x_b)$ for the random realization, as a function of $\zeta$. Notice that the size of this quantity decreases for $\zeta \to 0$ and $\zeta\to\pi$. 
  \small {\bf Right}: Black curve: Hellings Downs distribution. Blue points: response function for our random realization of pulsars, including the
  circular polarization kinematic dipole, with $\beta n_I=1/10$.  Notice that, for intermediate values of $\zeta$, the points are scattered
  with respect to the black line. 
   } 
  \label{fig_confinfA1bc}
\end{figure}

  In Figure \ref{fig_confinfA1bc}, left panel, we represent $\hat v\cdot (\hat x_a\times \hat x_b)$
 as a function of $\zeta$, for a random set of pulsar pairs whose positional vectors lie on a plane  orthogonal with the velocity vector $\hat v$. The size of the quantity  $\hat v\cdot (\hat x_a\times \hat x_b)$ is maximal for intermediate
 values of $\zeta\simeq\pi/2$, while it reduces at the extremes $\zeta\simeq0$ and $\zeta\simeq\pi$. In the right panel of the figure we represent the response
 function  $\Gamma^V_{ab}$ to circular polarization, which shows that scattering
 effects are more pronounced for intermediate values of the angle $\zeta$. 
 
Hence, to be maximally sensitive to circular
polarization $V$, it is convenient to focus on pulsars whose position vectors form a plane orthogonal to the direction of the velocity
vector $\hat v$, so that we completely remove possible
contaminations from kinematic effects due to GW intensity. 
The formula for the optimal signal-to-noise ratio ${\rm SNR}$ for detecting the presence of
$V$ in the PTA time-delay correlators is derived in Appendix \ref{AppB}.
It results 
 \bea
{\rm SNR}_V&=&2 \pi \,T^{1/2}\,\left\{\sum_{ab}\,\int d f\,\frac{
\bar V^2(f)\,\left(\Gamma_{ab}^V (f)\right)^2
}{S_a^{(n)}(f) \,S_b^{(n)}(f) }
\right\}^{1/2}\,,\nonumber
\\
&=&
2 \pi \,T^{1/2}\,\beta\,\left\{\sum_{ab}\,\int d f\,\frac{
\bar V^2(f)\,n_V^2(f)}{S_a^{(n)}(f) \,S_b^{(n)}(f) }  \left(\frac13+\frac{y_{ab} \ln y_{ab}}{4(1-y_{ab})} \right)^2\,\left[ \hat v\cdot (\hat x_a\times \hat x_b)\right]^2
\right\}^{1/2}
\label{opsnrv1}
\eea
where the sum is limited to pulsar pairs orthogonal to $\hat v$, as described above, and $S^{(n)}$
is the noise spectral density function (see Appendix  \ref{AppB} for more details). 
The general form for the circular polarization response $\Gamma_{ab}^V$ is given in eq \eqref{respva}, and
in the second line of eq \eqref{opsnrv1} we write its contributions up to the dipole. Besides
being proportional to $\beta$, it also depends on the slope of the quantity $\bar V(f)$, and might
be enhanced if  $\bar V(f)$ is a steep function of frequency. In fact, the study of kinematic
anisotropies -- given their unique dependence on the slope of the spectrum, see comments between eqs
 \eqref{defai2} and \eqref{defyab} --
might represent the only way to probe the frequency-dependence of quantities
associated with circular polarization.   In the most conservative cases,
  in case  $\bar V \simeq \bar I$
with no pronounced features,  in order to detect parity violating effects we would  need  a factor $1/\beta\simeq$ one thousand more in sensitivity 
with respect to current experiments.

\subsection{Kinematic anisotropies and SGWB  intensity }
\label{sec_str_in}

\begin{figure}
\begin{center}
        \includegraphics[width = 0.47 \textwidth]{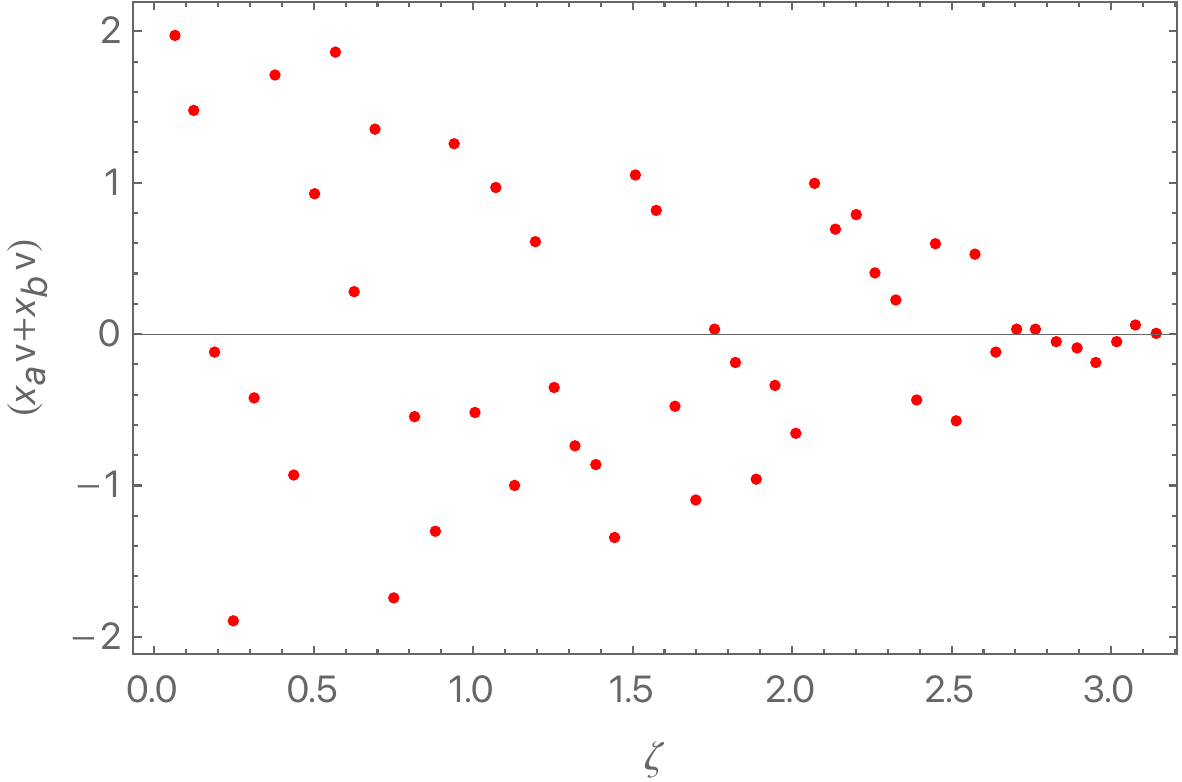}
    \includegraphics[width = 0.47 \textwidth]{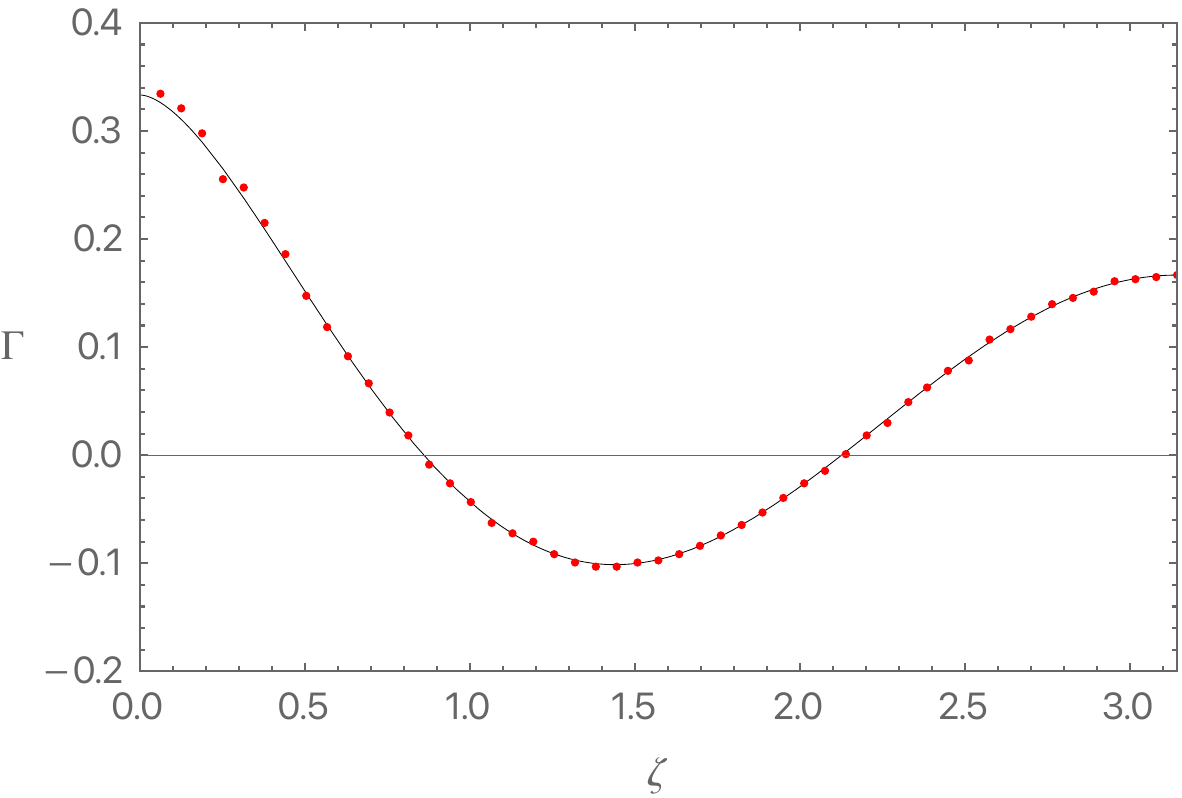}
    \end{center}
 \caption{\small We consider a random collection of pulsar pairs whose positional
 vectors $\hat x_a$ and $\hat x_b$ lie on the same plane as the relative velocity $\hat v$. Our aim is
 to represent how geometrical factors depending on $\hat v$ affect the response functions, as a function of the angle $\zeta$ among pulsars.
 {\bf Left } combination 
 $\hat v \cdot \hat x_a+ \hat v \cdot \hat x_b$ for the random realization, as a function of $\zeta$. Notice that the size of this quantity decreases as $\zeta\to\pi$. 
  \small {\bf Right}: Black curve: Hellings Downs distribution. Red points: response function for our random realization of pulsars, including the
  intensity kinematic dipole, with $\beta n_I=1/10$.  Notice that, for small values of $\zeta$, the points are scattered
  with respect to the black line. 
 } 
  \label{fig_confinfA1c1}
\end{figure}

Suppose  we are interested in  extracting from the kinematic anisotropy  contributions
associated with the GW intensity $\bar I(f)$ as introduced
in section \ref{sec_rin}. The corresponding  response
function $\Gamma^I_{ab}$
 of a pulsar
pair $(a,b)$ to intensity Doppler effects is given by 
eq \eqref{respia}, which we rewrite here up to the dipole contributions: 
\bea
\label{respia1a}
\Gamma^I_{ab}&=&
\left(\frac13-\frac{y_{ab}}{6}+y_{ab} \ln y_{ab} \right)
+{\beta\,n_I}  \left(\frac{1}{12}+\frac{ y_{ab}}{2}+\frac{ y_{ab} \ln y_{ab}}{2(1-y_{ab})} \right)
\, \left[\hat v\cdot \hat x_a+\hat v\cdot \hat x_b\right]\,,
\eea
Recall the definition of the parameter $y_{ab}=(1-\cos{\zeta})/2$ is expressed in terms of the angle  $\hat x_a\cdot \hat x_b = \cos \zeta$ between the unit vectors indicating the pulsar positions
with respect to the Earth. The function \eqref{respia1a} depends
on $\zeta$ through the first contribution between parenthesis, which corresponds
to   the Hellings Downs curve $\Gamma_{ab}^{\rm HD}$
as defined in eq \eqref{defFHD}. But eq  \eqref{respia1a}  also depends on the angle that each pulsar vector
forms  with the relative frame velocity  $\hat v$. As explained
 in section \ref{secptar},  only  pulsars whose vector components lie along $\hat v$  are sensitive to kinematic
anisotropies relative to the GW intensity. 

When plotting the quantity $\Gamma^I_{ab}$ as a function of $\zeta$
for each pulsar pair in a given set,   data points get
scattered around the Hellings-Downs curve of eq \eqref{defFHD} (see \cite{Taylor:2013esa,Mingarelli:2013dsa}). In fact, data  are modulated by the quantity
$(\hat v\cdot \hat x_a+\hat v\cdot \hat x_b)$ in eq \eqref{respia1a}; we expect that
  point scattering around the Hellings-Downs line 
 is maximal for small values of $\zeta$,
while it reduces for $\zeta\simeq\pi$, since in this limit the coefficient of the kinematic
dipole contribution vanishes: $(\hat v\cdot \hat x_a+\hat v\cdot \hat x_b)\simeq0$. We represent in Fig \ref{fig_confinfA1c1}, left panel, the value of the quantity  $(\hat v\cdot \hat x_a+\hat v\cdot \hat x_b)$ as a function of $\zeta$, for a random set of pulsars whose positions are coplanar with the velocity vector $\hat v$. As anticipated, 
the size of this quantity reduces as we increase the angle $\zeta$. The right panel
of the same figure shows the resulting response function $\Gamma^I_{ab}$ as function
of $\zeta$, which is indeed scattered for small angular separations $\zeta$
with respect to the Hellings Downs curve. Hence, the effects of kinematic anisotropies
 relative to the GW intensity $I$ are maximal for pulsar pairs lying in the same plane as $\hat v$, and
 with whose positional vectors form a small relative angle among themselves. This represents
 an interesting difference with respect to the GW circular polarization response that we discussed
in section \ref{sec_str_cp}.

\smallskip
As manifest from Fig \ref{fig_confinfA1c1},
the kinematic effects we are interested in are small. On the other hand,  by  combining signals \footnote{See the review \cite{Romano:2016dpx} for a discussion of possibilities  to combine GW signals to extract specific
physical information from them.}, 
we can form 
 null tests
for
 Doppler effects associated to the intensity of the SGWB. To find such
 combinations, we further exploit the results of  section \ref{secptar}. Suppose that GW are characterized by spin-2 polarization only, with no effects of parity violation. Then, 
  consider two pairs of pulsars. One pair,
$ab$, lies on a plane parallel to $\hat v$: hence it feels the effects of  kinematic anisotropies,
and the response function up to the dipole is given by eq \eqref{respia1a}. The 
other pair, $cd$, lies on a plane orthogonal to $\hat v$: hence it is blind to Doppler effects, and the pulsar response function is determined only   by the Hellings Downs curve of eq \eqref{defFHD}. 

We then take   the following combination  of equal-time correlators for time residuals (recall
their definition in eq \eqref{twopcordt}): 
\bea
\frac{
\left( \Gamma_{cd}^{\rm (HD)}  \right)\langle R_a(t) R_b(t) \rangle-
\left(\Gamma_{ab}^{\rm (HD)} \right) \langle R_c(t) R_d(t) \rangle}{\langle R_c(t) R_d(t) \rangle}&=&
\beta\,
\frac{\,F_{ab}^{(1)}
\int {d f\,n_I(f)\,\bar I(f)\, \sin^2{\left( \pi f t \right)}}/{f^2}}{\int {d f\,\bar I(f)\, \sin^2{\left( \pi f t \right)}}/{f^2}}
\,,
\label{difGA}
\eea
where $\Gamma^{\rm HD}$ is the Hellings Downs function of eq \eqref{defFHD},
while $F^{(1)}$ is given in eq \eqref{defF11}. Hence, in the context we are considering,  
 a measurement of  the
combination
\eqref{difGA} allows us to extract only the effects of kinematic anisotropies, with no
contaminations from the isotropic background.  Interestingly, given the dependence
on the intensity slope $n_I$ in the numerator of eq \eqref{difGA}, the sensitivity to kinematic anisotropies
get enhanced for signals with features, see for example section \ref{sec_someex}.  In this sense, kinematic anisotropies allow us to probe features in the frequency dependence
of GW correlators.

\begin{figure}
\begin{center}
    \includegraphics[width = 0.47 \textwidth]{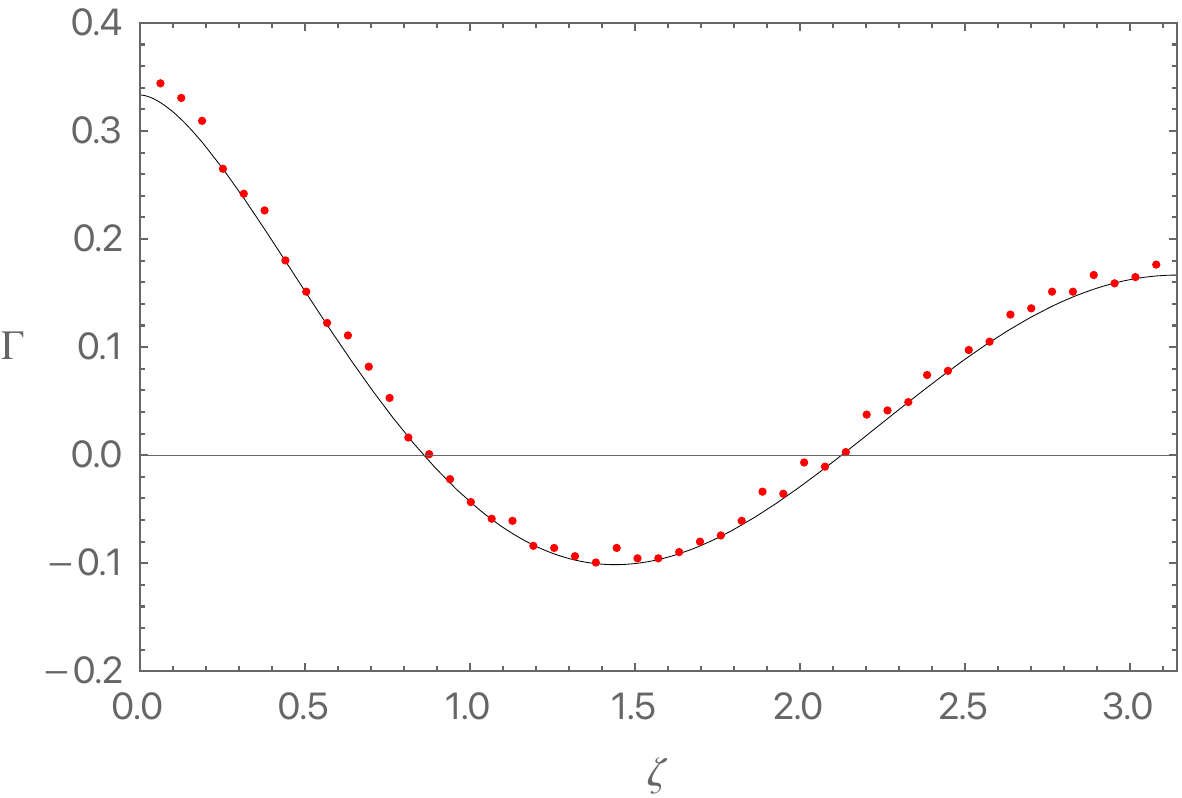}
        \includegraphics[width = 0.47 \textwidth]{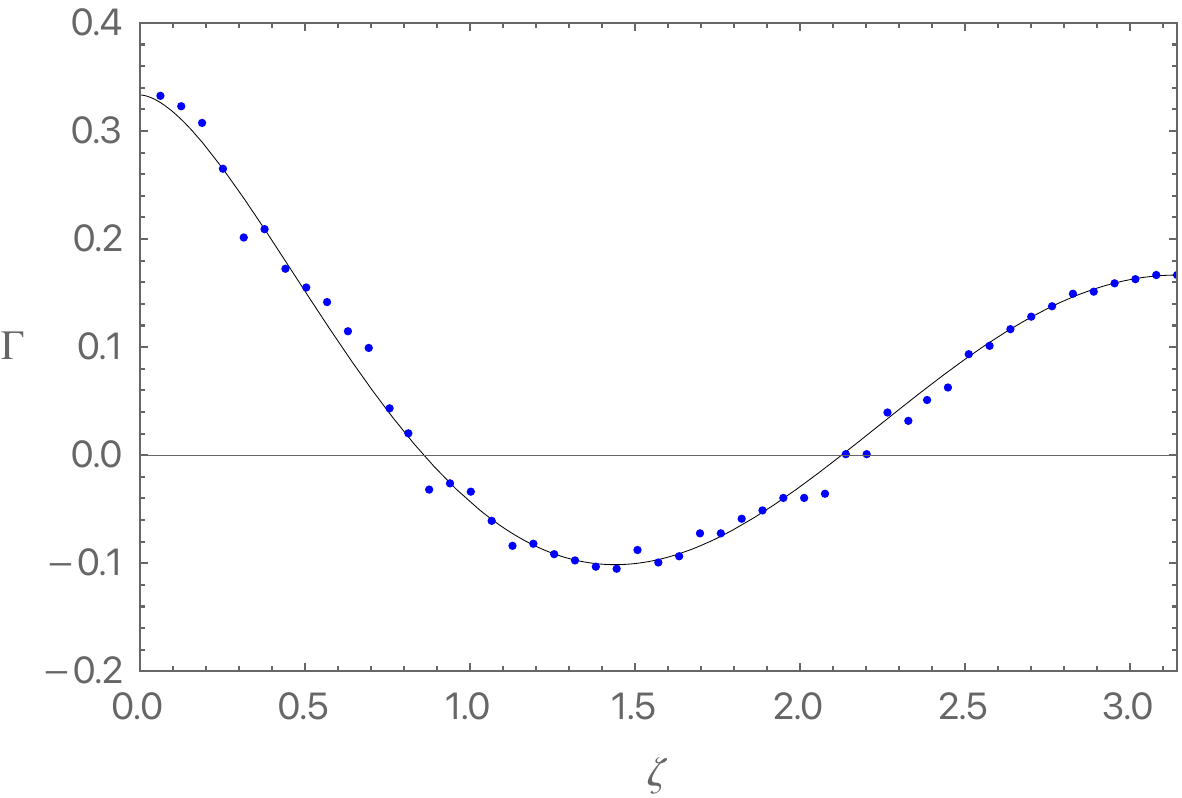}
    \end{center}
 \caption{ \small {
 This plot complements Figs \ref{fig_confinfA1c1} and \ref{fig_confinfA1bc}, by showing the
 contributions of the kinematic quadrupole to the angular dependence
 of the PTA response functions. Black lines: Hellings Downs curve.
 \bf Left panel:} The red points are  a random realisation
 of the  quadrupole contribution to the intensity response function, choosing the value $\beta^2 (\alpha_{\rm I}+n_{\rm I}^2- n_{\rm I}) =1/50$. {\bf Right panel:} The blue points are a random realization
 of the  quadrupole contribution to the    circular polarization response function, with $\beta^2 (\alpha_{\rm V}+n_{\rm V}^2- n_{\rm V})=1/10$.
 } 
  \label{fig_confinfA1aa31}
\end{figure}

\bigskip
 While so far we discussed the dipole only, similar considerations hold for the 
quadrupole contributions, that as we have seen can be enhanced in scenarios with features
in the isotropic SGWB as function of frequency. See Fig \ref{fig_confinfA1aa31} for representative
examples of kinematic quadrupole effects.

\subsection{Kinematic anisotropies and  scalar contributions to the GW signal}

Models of modified gravity often predict the existence of  light scalar fields associated with
  gravitational interactions. They 
contribute to GW  as additional spin-0, scalar polarizations \cite{Eardley:1973br}. Scalars can contribute
to the total GW intensity. But if their amplitude $\bar I_{\rm sc}$ is well smaller than the tensor intensity amplitude $\bar I_{\rm tn}$, it
might be  hard to disentangle  their contribution from analysing only  the isotropic part of the
background.

\smallskip

Interestingly, if the isotropic  scalar intensity $\bar I_{\rm sc}(f)$  has  a steep slope, or more generally it has pronounced 
 features associated with its   frequency dependence, its consequences on kinematic anisotropies can allow
us to extract the presence of scalar polarization with PTA data. 

\smallskip

In scenarios including contributions from the scalar polarization, we  express the total GW intensity (tensor plus scalar parts) as 
\be
\bar I(f)\,=\,\bar I_{\rm tn}+\bar I_{\rm sc}(f)\,.
\ee
Defining the intensity
 slope parameters as in eqs \eqref{defnsc}, \eqref{defasc}, 
 we find that the PTA intensity  response function to Doppler dipolar anisotropies
in such scalar-tensor theories can be decomposed as 
\bea
\label{respia1ab}
\Gamma^I_{ab}&=&\Gamma_{ab}^{\rm ST,\, iso}
+\frac{\beta}{12}\,\Gamma_{ab}^{\rm ST,\, dip}
\label{mixres1}
\eea
with
\bea
\Gamma_{ab}^{\rm ST,\, iso}&=&\frac{\bar I_{\rm tn}(f)}{\bar I_{\rm tn}+\bar I_{\rm sc}(f)}
\left(\frac13-\frac{y_{ab}}{6}+y_{ab} \ln y_{ab} \right)
+\frac{\bar I_{\rm sc}(f)}{\bar I_{\rm tn}+\bar I_{\rm sc}(f)}
\left(\frac13-\frac{y_{ab}}{6}
\right)
\\
\Gamma_{ab}^{\rm ST,\, dip}&=&\left[n_{\rm sc}+ {\,n_{\rm tn}}  \left({1}+{6\, y_{ab}}+\frac{ 6\,y_{ab} \ln y_{ab}}{(1-y_{ab})} \right) \right]
\, \left[\hat v\cdot \hat x_a+\hat v\cdot \hat x_b\right]\,.
\eea

The quantity $\Gamma_{ab}^{\rm ST,\, iso}$ describes the pulsar response to the isotropic
part of the scalar-tensor intensity, while $\Gamma_{ab}^{\rm ST,\, dip}$   the pulsar response to the dipolar
kinematic anisotropy. 
Hence, even if $\bar I_{\rm sc}/\bar I_{\rm tn}\ll1$  -- so that $\Gamma_{ab}^{\rm ST,\, iso}$   is `almost blind' to scalar contributions --  a pronounced   slope dependence, leading to $n_{\rm sc}\ge n_{\rm tn}$
 can make   $\Gamma_{ab}^{\rm ST,\, dip}$    sensitive to the scalar polarization. An appropriate combination of signals associated to different pulsar pairs, following the method described
 around eq \eqref{difGA},  isolates the effects of scalar dipolar anisotropies in the response functions. Consequently, the presence of scalar polarizations 
 might be detectable even if scalar polarizations contribute as  small part to the GW signal amplitude.

\section{Outlook}

The recent detection of a stochastic gravitational wave background
from several pulsar timing array collaborations opens a new window
for gravitational wave cosmology. If the stochastic gravitational
wave background has a cosmological origin, one of its guaranteed
features are kinematic anisotropies due to our motion with respect 
to the gravitational wave source. In this work, we investigated the
physical information that can be extracted from a detection of kinematic
anisotropies. We shown that the pulsar response functions
to Doppler effects depend on the stochastic background source and on the underlying 
gravitational  physics.  Furthermore, we stressed for the first time how the measured
gravitational wave 
signal depends on the pulsar location with respect to the relative velocity among frames. We presented our results in a convenient way that emphasizes the geometrical
dependence of the pulsar response functions to different properties of the stochastic background. 

\smallskip

Our findings can be useful in planning future measurements, or in elaborating existing data, for using 
kinematic anisotropies to detect, or set more
stringent constraints, on parity violation in gravitational interactions, on 
the presence of scalar/vector  polarizations, or to develop independent
probes of the frequency profile of the stochastic background.    
Besides
backgrounds detectable at nano-Hertz frequencies with pulsar timing arrays,
it would be interesting to apply our findings also to astrometric
measurements of gravitational waves based on Gaia  data, see e.g. the discussion in \cite{Book:2010pf,Moore:2017ity,Mihaylov:2018uqm,OBeirne:2018slh,Qin:2018yhy}. 

\smallskip

Current measurements (see  \cite{NANOGrav:2023tcn}) set upper bounds $C_{\ell>0} / C_0 \le 0.2$  on the amplitude of multipoles of the SGWB.  The work  \cite{Hotinli:2019tpc} estimates  that a detection is possible only for anisotropies whose size is of the order of 0.1 with respect to the isotropic background, unless the amplitude of the latter is measured with high SNR. 
However,  the recent first detection of the isotropic part of the background by various international PTA collaborations gives hope for the future. Future measurements of pulsar timing residuals, for example  by the SKA collaboration, will monitor many more pulsars than current experiments. Their higher sensitivity will improve the confidence on the detection of the isotropic part of the SGWB and its  SNR, enhancing the opportunity to detect SGWB anisotropies.

The theoretical study in this manuscript adds further ingredients that can help in planning future dedicated searches of Doppler anisotropies. 
Moreover,  the manuscript studies the consequences of SGWB with a frequency dependence that goes beyond a  power law,  as motivated for example by early universe models producing primordial black holes (see section \ref{sec_someex}). In this case, we cannot apply the factorizable Ansatz for the anisotropic SGWB often used in the literature, as constituted of a  part depending on frequency, times a part depending on direction.  In fact, we find that a SGWB with enhanced features in its frequency dependence can amplify the size of kinematic anisotropies to values well larger than the amplitude $v/c\sim 10^{-3}$ one would expect for kinematic anisotropies.

\smallskip

In our work, we assumed that the direction and amplitude of the velocity vector among the cosmological
source of gravitational waves and the solar sytem baricenter coincides to what measured
by the cosmic microwave background dipole. On the other hand, our formulas can be used
to directly test this hypothesis,  somehow analogously to  the proposal  \cite{10.1093/mnras/206.2.377} in the different context of radio sources. In fact, the topic of radio galaxy and quasar measurements of the cosmic dipole is a point of debate
in recent literature  \cite{Blake:2002gx,Singal:2011dy,Gibelyou:2012ri,Rubart:2013tx,Secrest:2020has,Secrest:2022uvx}. It would be  interesting to use gravitational waves as an independent probe of the direction of the kinematic dipole, and  test alternative
scenarios based on large scale superhorizon isocurvature fluctuations  \cite{Turner:1991dn,Langlois:1995ca},   intrinsic anisotropies \cite{Roldan:2016ayx}, or more general deviations from the cosmological principle \cite{Aluri:2022hzs}. In this sense, this program belongs to the `multimessenger cosmology' framework advocated in \cite{Adshead:2020bji,Ricciardone:2021kel}.

\subsection*{Acknowledgments}

It is a pleasure to thank Debika Chowdhury, Ameek Malhotra (especially), Maria Mylova, and Ivonne Zavala for useful input. GT is partially funded by the STFC grant ST/T000813/1. For the purpose of open access, the author has applied a Creative Commons Attribution licence to any Author Accepted Manuscript version arising.

\begin{appendix}

\section{Identities involving the polarization tensors}
\label{AppA}

We derive useful identities involving the polarization
tensors entering in the Fourier transform of the GW signals,
which we used in sections \ref{secsetup}-\ref{sec_MG}.
We start
discussing the spin-2 case, to then consider respectively
the spin-0 and spin-1 cases.

\medskip
\noindent
{\bf Spin-2 polarization tensors:} 
In the main text, section \ref{secptar}, we focussed on the $(+,\times)$ basis for the polarization tensors. 
To develop our arguments,  in this Appendix it is convenient to introduce the 
 $(R,L)$ circular basis:
 \be
 \label{relasc1}
{\bf e}_{ab}^{R/L}(\hat n)\,=\,\frac{{\bf e}_{ab}^{+}(\hat n) \pm\,i \,{\bf e}_{ab}^{\times}(\hat n) }{\sqrt 2} \,.
\ee
We assume that the polarization tensors ${\bf e}_{ab}^{+,\times}$
 are real quantities: hence $
 \left({\bf e}_{ab}^{R}\right)^*
 \,=\,{\bf e}_{ab}^{L}
 $. 
Notice also the identity 
 \bea
{\bf e}_{ab}^{R}(\hat n)
\,{\bf e}_{cd}^{L}(\hat n)&=&\frac12 
\left( 
{\bf e}_{ab}^{+} {\bf e}_{cd}^{+}
+
{\bf e}_{ab}^{\times} {\bf e}_{cd}^{\times}
\right)
+\frac{i}{2}
\left( 
{\bf e}_{ab}^{\times} {\bf e}_{cd}^{+}
-
{\bf e}_{ab}^{+} {\bf e}_{cd}^{\times}
\right)\,.
\label{sumpol1}
\eea

 We identify the circular polarization indexes with $\pm1$: $\lambda\,=\,R/L\,=\,\pm1$. 
 The polarization tensors in the circular basis can be conveniently expressed as
 a  product of polarization vectors:
\be
\label{scomp1}
{\bf e}_{ab}^{\lambda}(\hat n)\,=\,{\bf e}_{a}^{\lambda}(\hat n){\bf e}_{b}^{\lambda}(\hat n)
\,=\,
\frac{\hat p_a+i \lambda \hat q_a}{\sqrt{2}}\,\frac{\hat p_b+i \lambda \hat q_b}{\sqrt{2}}
\,.
\ee 
The real unit vectors $\hat p_a$, $\hat q_a$ are orthogonal among themselves and to 
the GW direction $\hat n$. We can express $\hat q$ as 
\bea
\hat q_a&=&\epsilon_{abc} \hat n_b  \hat p_c\,,
\eea
with $\epsilon_{abc}$ the Levi-Civita tensor in three dimensions. (For what comes next, we do not need to make any specific choice for the direction $\hat p_a$.)
It is easy to verify the following identity involving symmetric tensors
\be \label{reldot1}
\hat p_a \hat p_b+\hat q_a \hat q_b\,=\,\delta_{ab}-\hat n_a \hat n_b
\,.
\ee
In fact,  both the LHS and RHS are parallel to $\hat p_a$ and $\hat q_a$, orthogonal to $\hat n_a$, and of trace 2. Moreover, the antisymmetric combination $\hat p_a \hat q_b-\hat p_b \hat q_a$ is orthogonal to $\hat n_a$, and satisfy the identity
\be\label{relcros1}
\hat p_a \hat q_b-\hat p_b \hat q_a\,=\,\epsilon_{abc} \,\hat n_c\,.
\ee
We can convince ourselves of the validity of  relation \eqref{relcros1} by contracting with 
$\hat p_a$, $\hat q_b$:
\be
\sum_a \hat p_a (\hat p_a \hat q_b-\hat p_b \hat q_a)
\,=\,\hat q_b\,=\,\hat p_a \epsilon_{abc} \,\hat n_c\,,
\ee
and
\be
\sum_a 
\hat q_a (\hat p_a \hat q_b-\hat p_b \hat q_a)\,=\,-\hat p_b\,=\,\hat q_a \epsilon_{abc} \,\hat n_c\,,
\ee
compatibly with equation \eqref{relcros1}. See also \cite{Domcke:2019zls}.

 Identities \eqref{reldot1} and \eqref{relcros1} ensure the relation (no summation
 over polarization indexes)
\bea
{\bf e}_{a}^{\lambda}(\hat n) \left({\bf e}_{b}^{\lambda}(\hat n)\right)^*
&=&\frac{p_a+i \lambda q_a}{\sqrt{2}}\,\frac{p_b-i \lambda q_b}{\sqrt{2}}\,,
\\
&=&\frac12 \left( p_a p_b+q_a q_b \right)-\frac{i \lambda}{2}
\left(p_a q_b-p_b q_a\right) \,,
\\
&=&\frac12\left(\delta_{ab}-n_a n_b\right)-\frac{i \lambda}{2}\epsilon_{abc} \,n_c\,.
\eea
Hence:
\bea
{\bf e}_{ab}^{\lambda}(\hat n)({\bf e}_{cd}^{\lambda}(\hat n))^*
&=&{\bf e}_{a}^{\lambda}(\hat n) \left({\bf e}_{c}^{\lambda}(\hat n)\right)^*
{\bf e}_{b}^{\lambda}(\hat n) \left({\bf e}_{d}^{\lambda}(\hat n)\right)^*
\\
&=&\frac14 
\left(\delta_{ac}-n_a n_c-{i \lambda}\epsilon_{acf} \,n_f\right)
\left(\delta_{bd}-n_b n_d-{i \lambda}\epsilon_{bdg} \,n_g\right)
\,.
\eea
Expanding the product, and passing from the circular $(R,L)$
polarization to the original $(+,\times)$ polarization  using
eq \eqref{sumpol1},  we find the
identities:
\bea
\sum_{\lambda \lambda'}
{\bf e}_{ab}^\lambda (\hat n) {\bf e}_{cd}^{\lambda'} (\hat n) 
 \delta_{\lambda \lambda'}&=&\frac12 (\delta_{ac}-n_a n_c)
  (\delta_{bd}-n_b n_d)+\frac12 (\delta_{ad}-n_a n_d)
  (\delta_{bc}-n_b n_c)
  \nonumber
  \\
  &&-\frac12 (\delta_{ab}-n_a n_b)
  (\delta_{cd}-n_c n_d)\,,
  \label{impeq1}
\eea
and 
\bea
\label{impeq2}
\sum_{\lambda \lambda'}
{\bf e}_{ab}^\lambda (\hat n) {\bf e}_{cd}^{\lambda'} (\hat n) 
 \epsilon_{\lambda \lambda'}&=&\frac12 (\delta_{ac}-n_a n_c) \epsilon_{bdf} n_f
 +\frac12 (\delta_{bd}-n_b n_d) \epsilon_{acf} n_f\,.
 \eea
 We used these relations in the main text to obtain eqs \eqref{contra} and \eqref{contrb},
 controlling
 the PTA response respectively to GW intensity and circular
polarization.

\medskip
\noindent
{\bf Spin-0 polarization tensor:}
We can be brief here, since the spin-0 mode is associated with a single
polarization tensor
\be
\sigma_{ab}(\hat n)\,=\,\hat n_a \hat n_b\,,
\ee
which we used for deriving equation \eqref{GammaSc1} in the main text. 

\medskip
\noindent

\medskip
\noindent
{\bf Spin-1 polarization tensors:} 
To describe  spin-1 polarization tensors we can
use  a  $(+,\times)$ basis or a circular $(R,L)$ basis: they are related
exactly as in eq \eqref{relasc1} for the spin-2 case. 
In the circular basis $(R,L)\,=\,\pm1$, the vector polarization tensors read
\be
{\bf u}_{ab}^{\lambda}(\hat n)\,=\,\frac{1}{ \sqrt 2} 
\left( {\bf e}_{a}^{\lambda}(\hat n) \hat n_b+
\hat n_a\,{\bf e}_{b}^{\lambda}(\hat n)
\right)\,,
\ee
with $ {\bf e}_{a}^{\lambda}(\hat n) $ the vectors appearing in eq \eqref{scomp1}.
We have the relation
\bea
{\bf u}_{ab}^{\lambda}(\hat n)\,
\left({\bf u}_{cd}^{\lambda}(\hat n)\right)^*
&=&\frac12 \left({\bf e}_{a}^{\lambda}(\hat n) \hat n_b+
\hat n_a\,{\bf e}_{b}^{\lambda}(\hat n) \right) 
 \left( \left( {\bf e}_{c}^{\lambda}(\hat n) \right)^* \hat n_d+
\hat n_c\,
\left( {\bf e}_{d}^{\lambda}(\hat n) \right)^*
\right) \,,
\\
&=&
\frac12 \Big[
{\bf e}_{a}^{\lambda}(\hat n)
 \left( {\bf e}_{c}^{\lambda}(\hat n) \right)^* \hat n_b \hat n_d
 +
{\bf e}_{b}^{\lambda}(\hat n)
 \left( {\bf e}_{c}^{\lambda}(\hat n) \right)^* \hat n_a \hat n_d
 \nonumber
 \\
 &&
 +{\bf e}_{a}^{\lambda}(\hat n)
 \left( {\bf e}_{d}^{\lambda}(\hat n) \right)^* \hat n_b \hat n_c
 +
{\bf e}_{b}^{\sigma}(\hat n)
 \left( {\bf e}_{d}^{\lambda}(\hat n) \right)^* \hat n_a \hat n_c
\Big]\,,
\\
&=&
\frac14\left[\delta_{ac}-\hat n_a \hat n_c-{i \lambda}\epsilon_{acf} \,n_f\right]
\hat n_b \hat n_d 
\nonumber
\\
&+&
\frac14\left[\delta_{bc}-\hat n_b \hat n_c-{i \lambda}\epsilon_{bcf} \,\hat n_f\right]
\hat n_a \hat n_d \,,
\nonumber
\\
&=&\frac14 \left( \delta_{ac} \hat n_b\hat  n_d
+\delta_{bc} \hat n_a \hat n_d+ \delta_{ad} \hat n_b \hat n_c+
\delta_{bd}\hat  n_a \hat n_c
\right)-\hat n_a\hat n_b \hat n_c  \hat n_d
\nonumber
\\&-&\frac{i \lambda}{4}
\left( \epsilon_{acf}  \hat n_b \hat n_d+\epsilon_{bcf}  \hat n_a\hat n_d
+\epsilon_{adf}  \hat n_b \hat n_c+\epsilon_{bdf}  \hat n_a \hat n_c
\right)\hat n_f\,.
\eea
Proceeding  as we did in deriving eqs \eqref{impeq1} and \eqref{impeq2} for the spin-2 case, we can then obtain the
following identities for vector polarizations in the $(+,\times)$ basis:
\bea
\sum_{\lambda \lambda'}
{\bf u}_{ab}^\lambda (\hat n) {\bf u}_{cd}^{\lambda'} (\hat n) 
 \delta_{\lambda \lambda'}&=&
 \frac12 \left( \delta_{ac} \hat n_b\hat  n_d
+\delta_{bc} \hat n_a \hat n_d+ \delta_{ad} \hat n_b \hat n_c+
\delta_{bd}\hat  n_a \hat n_c
\right)-2 \hat n_a\hat n_b \hat n_c  \hat n_d\,,
  \label{impeq1v}
\eea
and 
\bea
\label{impeq2v}
\sum_{\lambda \lambda'}
{\bf u}_{ab}^\lambda (\hat n) {\bf u}_{cd}^{\lambda'} (\hat n) 
 \epsilon_{\lambda \lambda'}&=&\frac12
 \left( \epsilon_{acf}  \hat n_b \hat n_d+\epsilon_{bcf}  \hat n_a\hat n_d
+\epsilon_{adf}  \hat n_b \hat n_c+\epsilon_{bdf}  \hat n_a \hat n_c
\right)\hat n_f\,,
 \eea
that we used for deriving equation \eqref{GammaVe1} in the main text.

\section{The optimal signal-to-noise ratio}
\label{AppB}

In this Appendix we discuss the computation of the signal-to-noise ratio, in terms
of a match-filtering technique as discussed in the paper \cite{Anholm:2008wy} and the textbook \cite{Maggiore:2007ulw}. 
We focus here on the case of spin-2
polarizations, but the arguments are the same for spin-1 and spin-0. 
We assume
that the noise dominates over the signal. The  latter can be extracted once we
know its properties. 

The stationary two-point correlator for time-delays
of pulsar measurements
summed over a PTA set of pulsar pairs $(a,b)$ is
weighted by an appropriate choice of filters
\be
{\cal Y}\,=\,\sum_{ab}\,\int_{-T/2}^{T/2} d t_1 \int_{-T/2}^{T/2} d t_2\,z_a(t_1) z_b(t_2)
\,Q_{ab}(t_1-t_2)\,,
\ee
where $T$ indicates the duration of the experiment. 
We pass to Fourier space, and write the previous expression as
\be
{\cal Y}\,=\,\frac12 \sum_{ab}\,\int_{-\infty}^{\infty} d f_1 d f_2\,\delta_{T}(f_1-f_2)
\,z_a^*(f_1)z_b(f_2)\,Q_{ab}(f_2) +c.c.
\ee
The `finite-time' delta-function is
\be
\delta_T(f)\,=\,\frac{\sin{(\pi f T) }}{\pi f}
\ee
We assume that $Q_{ab}(t)$ is real, and its Fourier transform obeys $Q_{ab}^*(f)\,=\,Q_{ab}(-f)$. Decomposed in real and imaginary part, we can express it as
  \be
  \label{qdec}
  Q_{ab}(f)\,=\,Q_{ab}^{I}(f)+i\,Q_{ab}^{V}(f)\,,
  \ee
with $Q_{ab}^{I,V}$ two real functions of frequency. 
In defining the SNR, the signal $S$
is the averaged value of ${\cal Y}$
when the GW signal is present; the noise $N$
is root mean square of ${\cal Y}$ when the GW is absent.
The aim is to determine the optimal filter functions $Q^{I}$ and
$Q^{V}$ for extracting the GW signal
from the noise. 

\smallskip
We start expressing the signal $S$, as
\bea
S\,=\,\langle {\cal Y} \rangle\,=\,\frac12 \sum_{ab}\,\int_{-\infty}^{\infty} d f_1 d f_2\,\delta_{T}(f_1-f_2)
\,\langle z_a^*(f_1)z_b(f_2) \rangle\,Q_{ab}(f_2) +c.c.
\label{defSIG}
\eea
Using formulas developed in section \ref{secsetup},
we can
write the two-point correlator in Fourier space as 
\bea
\langle z_a^*(f_1)z_b(f_2) \rangle&=&\frac12
\int d^2 \hat n\,D_a^\lambda D_b^{\lambda'}
\left( I(f_1,\hat n)\,\delta_{\lambda \lambda'}-i V(f_1,\hat n)\,\epsilon_{\lambda \lambda'} 
\right)\,\delta(f_1-f_2)
\\
&=&\pi\,\bar{I}(f_1)\,\left( \Gamma_{ab}^I (f_1)-  i
\, \Gamma_{ab}^V (f_1) \right)\,.
\eea
We plug this expression in eq \eqref{defSIG},  and use the decomposition
\eqref{qdec}, finding
\be
S\,=\,\pi\,T\,\sum_{ab}\,\int df\,\bar I(f)\,\left[ \Gamma_{ab}^I (f)
Q_{ab}^{I}(f)+ \Gamma_{ab}^V (f)
Q_{ab}^{V}(f)
\right]
\ee
since $\delta_T(0)=T$. 
The computation of the noise $N$ is identical to the textbook 
treatment of \cite{Maggiore:2007ulw}. Characterizing the  correlator of pulsar noise $n_a$ in 
terms of the quantity $S_a^{(n)}$ as
\be
\langle n_a^*(f_1) n_a(f_2)\rangle\,=\,\frac12 \delta(f_1-f_2)\,S_a^{(n)}(f_1)\,,
\ee
the square of the noise is
\be
N^2\,=\,\frac{T}{4}\,\sum_{ab}\int d f\,Q_{ab}(f) Q_{ab}^*(f)  \,S_a^{(n)}(f) \,S_b^{(n)}(f)
\,.
\ee
Hence, the signal-to-noise ratio results
\bea
{\rm SNR}\,=\,T^{1/2}\,\frac{2 \pi\,
\sum_{ab}\,\int df\,\bar I(f)\,\left[ \Gamma_{ab}^I (f)
Q_{ab}^{I}(f)+ \Gamma_{ab}^V (f)
Q_{ab}^{V}(f)
\right]
}{\left[ \sum_{cd}\int d f\,\left[ (Q_{cd}^{I}(f))^2
+(Q_{cd}^{V}(f))^2
\right]  \,S_c^{(n)}(f) \,S_d^{(n)}(f)
\right]^{1/2}}\,.
\label{snre1}
\eea
We now determine the optimal choice of filter function
that maximises the SNR. Given a tensor  with structure
\be
({\cal A}^{(1)}, {\cal A}^{(2)})\,=\,\left( (A_{ab}^{(1)},A_{ab}^{(2)} ),  (A_{cd}^{(1)},A_{cd}^{(2)} )
\dots\right)
\ee 
where the dots run over the pulsar pairs, one considers the positive-definite product
\be
\big\langle ({\cal A}^{(1)}, {\cal A}^{(2)}), ({\cal B}^{(1)}, {\cal B}^{(2)}) \big\rangle\,=\,\sum_{ab}
\,\int d f \left( A_{ab}^{(1)} A_{ab}^{(2)}+B_{ab}^{(1)} B_{ab}^{(2)}  \right)S_a^{(n)}(f) \,S_b^{(n)}(f)
\ee
Using this tool,
we can schematically express the SNR of eq \eqref{snre1} as 
\be
\label{scalSNR1}
{\rm SNR}\,=\,2 \pi \,T^{1/2}\,\frac{\big\langle (Q^{(1)},Q^{(2)} ), (\bar I \,\Gamma^I/S_n^2 , \bar V \,\Gamma^V/S_n^2)\big\rangle}{
\big\langle (Q^{(1)},Q^{(2)} ), (Q^{(1)},Q^{(2)} ) \big\rangle \,.
}
\ee

\smallskip
\noindent
We now have different options:
\begin{itemize}
\item  If we wish to maximise the SNR to the total signal (intensity and circular polarization
combined) we choose the filter  $(\bar I \,\Gamma^I/S_n^2 , \bar V \,\Gamma^V/S_n^2)$
which obviously maximises the scalar product of eq \eqref{scalSNR1}. 
  Using this filter we
conclude that the optimal SNR is
\be
{\rm SNR}_{I+V}\,=\,2 \pi \,T^{1/2}\,\left\{\sum_{ab}\,\int d f\,\frac{
 \bar I^2(f)\,\left(\Gamma_{ab}^I (f)\right)^2+\bar V^2(f)\,\left(\Gamma_{ab}^V (f)\right)^2
}{S_a^{(n)}(f) \,S_b^{(n)}(f) }
\right\}^{1/2}\,.
\ee
Its value depends on the GW intensity and circular polarization, and on the geometric
quantities constituting  $\Gamma_{ab}^I$ and
$\Gamma_{ab}^V$, as discussed in  the main text.
\item We can instead be interested to extract  from the signal {\it only} components
associated for example to circular polarization, and determine the filter better suited
to this purpose. The filter is then $(0, \bar V \,\Gamma^V/S_n^2)$ (i.e. a filter
whose Fourier transform has only imaginary and no real component), and the associated
SNR is
\be
{\rm SNR}_V\,=\,2 \pi \,T^{1/2}\,\left\{\sum_{ab}\,\int d f\,\frac{
\bar V^2(f)\,\left(\Gamma_{ab}^V (f)\right)^2
}{S_a^{(n)}(f) \,S_b^{(n)}(f) }
\right\}^{1/2}\,.
\ee
As discussed in the main text, the sensitivity to circular polarization is then enhanced
by selecting pulsars whose directions form a plane orthogonal to the velocity vector
$\hat v$ between SGWB source and detector frames. 

\item
Analogously, if we wish only to be sensitive to the GW intensity, we select a filter
with real part only, and the maximal SNR is
\be
{\rm SNR}_I\,=\,2 \pi \,T^{1/2}\,\left\{\sum_{ab}\,\int d f\,\frac{
\bar I^2(f)\,\left(\Gamma_{ab}^I (f)\right)^2
}{S_a^{(n)}(f) \,S_b^{(n)}(f) }
\right\}^{1/2}\,.
\ee
The  sensitivity to intensity is  enhanced
by selecting pulsars whose directions are parallel to the velocity vector
$\hat v$ between SGWB source and detector frames. 

\end{itemize}
\end{appendix}

{\small
\providecommand{\href}[2]{#2}\begingroup\raggedright\endgroup



}

\end{document}